\documentclass[twocolumn]{article}

\usepackage{arxiv}

\usepackage[utf8]{inputenc} 
\usepackage[T1]{fontenc}    
\usepackage{hyperref}       
\usepackage{url}            
\usepackage{booktabs}       
\usepackage{amsfonts}       
\usepackage{nicefrac}       
\usepackage{microtype}      
\usepackage{graphicx}
\usepackage[numbers,sort&compress]{natbib}
\setcitestyle{comma,numbers,open={[},close={]}}
\usepackage{doi}

\title{Viscoelasticty with physics-augmented neural networks: Model formulation and training methods without prescribed internal variables}

\author{
	Max Rosenkranz\\
	Chair of Scientific Computing \\
    for	Systems Biology\\
	TU Dresden,
	01062 Dresden, Germany \\
	\And
	Karl A. Kalina\\
	Chair of Computational and\\
	Experimental Solid Mechanics\\
	TU Dresden,
	01062 Dresden, Germany \\
	\And
	J\"{o}rg Brummund\\
	Chair of Computational and\\
	Experimental Solid Mechanics\\
	TU Dresden,
	01062 Dresden, Germany \\
	\And
	WaiChing Sun\\
	Department of Civil Engineering\\
	and Engineering Mechanics\\
	Columbia University, 
	NY 10027, New York, United States \\
	\And
	Markus K\"{a}stner\thanks{Corresponding author, email: \texttt{markus.kaestner@tu-dresden.de}.} \\
	Chair of Computational and\\
	Experimental Solid Mechanics\\
	TU Dresden, 
	01062 Dresden, Germany \\
}

\renewcommand{\shorttitle}{Viscoelasticty with physics-augmented neural networks}

\hypersetup{
pdftitle={Preprint_RosenkranzEtAl_2024},
pdfauthor={Max Rosenkranz},
}

\input{StyleSetup}

\begin{document}

\twocolumn[
\begin{@twocolumnfalse}
\maketitle
\begin{abstract}
We present an approach for the data-driven modeling of nonlinear viscoelastic materials at small strains which is based on physics-augmented neural networks (NNs) and requires only stress and strain paths for training. The model is built on the concept of generalized standard materials and is therefore thermodynamically consistent by construction. It consists of a free energy and a dissipation potential, which can be either expressed by the components of their tensor arguments or by a suitable set of invariants. The two potentials are described by fully/partially input convex neural networks.
For training of the NN model by paths of stress and strain, an efficient and flexible training method based on a recurrent cell, particularly a long short-term memory cell, is developed to automatically generate the internal variable(s) during the training process. The proposed method is benchmarked and thoroughly compared with existing approaches.
These include a method that obtains the internal variable by integrating the evolution equation over the entire sequence, while the other method uses an  an auxiliary feedforward neural network for the internal variable(s). Databases for training are generated by using a conventional nonlinear viscoelastic reference model, where 3D and 2D plane strain data with either ideal or noisy stresses are generated. The coordinate-based and the invariant-based formulation are compared and the advantages of the latter are demonstrated. Afterwards, the invariant-based model is calibrated by applying the three training methods using ideal or noisy stress data. All methods yield good results, but differ in computation time and usability for large data sets. The presented training method based on a recurrent cell turns out to be particularly robust and widely applicable and thus represents a promising approach for the calibration of other types of models as well.
\end{abstract}
\end{@twocolumnfalse}
]


\section{Introduction}
\label{sec:Intro}
Despite a large number of existing classical constitutive models for nonlinear elastic and inelastic materials \cite{Haupt2000,deSouzaNeto2008a}, the description of novel materials with complex constitutive behavior remains a challenging task. The choice of a suitable model and the identification of the corresponding material parameters is time-consuming and does not necessarily lead to results with the desired accuracy, so that the development of new specialized models may be necessary. For this reason, a new class of approaches has emerged in recent years, which are often referred to as \textit{data-driven} or \emph{data-based} methods \cite{Bock2019,Dornheim2023}.
In particular, the application of \emph{artificial neural networks (NNs)} \cite{minsky_perceptrons_1972} has become popular and has the potential to replace conventional modeling strategies under certain circumstances. As universal function approximators, they are capable of expressing complex relations that are difficult to grasp with manual modeling efforts, given that the training is successful.

\subsection{Constitutive modeling with neural networks}
Historically, the pioneering work of Ghaboussi~et~al.~\cite{ghaboussi_knowledgebased_1991} from the beginning of the 1990s is particularly noteworthy. In this work, NNs, specifically \emph{feedforward neural networks (FNNs)}, are used for the first time to predict hysteresis in uniaxial and multiaxial stress states. Therein, the FNN is fed with information from several previous time steps to enable it to learn history-dependent behavior. In \citeA{furukawa_implicit_1998}, a constitutive model based on an FNN is presented that can be used to train uniaxial viscoplastic behavior. However, this requires internal variables in the training process, whose experimental identification is also described. Despite the rather simple nature of these models from today's point of view, approaches of this kind indicate the potential of data-driven constitutive modeling. Without having to decide on a specific model, it is possible to learn complex material behavior. With the recent rise in popularity of NNs and the associated rapid progress in efficiency and accessibility, many different methods have emerged in an extremely short time, extending and improving these approaches.
For example, the works \cite{hambli_multiscale_2011, yao_artificial_2014, lu_data-driven_2019, Heider2020, Fuchs2021} propose advanced techniques using FNNs, while \cite{ghavamian_accelerating_2019, mei_study_2020, Bonatti2022, Heider2020, Fuchs2021} make use of \emph{recurrent neural networks (RNNs)}, which are capable of making predictions based on past events due to their internal structure \cite{rumelhart_learning_1988}.

Thus, recurrent architectures are an appealing way to model inelastic behavior, since the provision of history variables or internal variables can be avoided \cite{rosenkranz_comparative_2023}. In particular, the development of advanced RNN cells such as \emph{long short-term memory (LSTM)} \cite{hochreiter_long_1997} or \emph{gated recurrent units (GRUs)} \cite{cho2014learning}, which provide increased memory capacity and enable an efficient training, lead to great popularity and rapid progress in this field.

What remains a challenge for NN-based approaches is the lack of a fundamental inclusion of physics, especially the \emph{second law of thermodynamics}. The networks map input quantities directly to the variable of interest, for example the stress, and thus approximate only the phenomenological relationship between input and output \cite{ghaboussi_knowledgebased_1991,furukawa_implicit_1998}. Compared to most classical constitutive models, the physical motivation is completely missing. This has some significant downsides. Such models, often denoted as \emph{black box models}, cannot guarantee robustness of the predictions beyond the training range covered by the used training data. Exceeding this range may not only lead to wrong but also to physically implausible predictions \cite{linden_neural_2023,masi_thermodynamics-based_2021}. Furthermore, the training process is entirely driven by the training data and is not supported by existing knowledge from classical constitutive modeling. This may lead to more complex optimization problems that exhibit gradient conflicts and require more training data \cite{masi_thermodynamics-based_2021}. For this reason, it seems natural to integrate the existing knowledge from continuum mechanics and constitutive modeling into the NN architecture to combine the advantages of both concepts.
This kind of NN-based constitutive models or scientific machine learning approaches in general are labeled as \emph{physics-informed} \cite{raissi_physics-informed_2019,henkes_physics_2022}, \emph{mechanics-informed} \cite{asad_mechanicsinformed_2022}, \emph{physics-augmented} \cite{klein_finite_2022}, \emph{physics-constrained} \cite{kalina_feann_2023}, \emph{thermodynamics-informed} \cite{Vlassis2021} or \emph{thermodynamics-based} \cite{masi_thermodynamics-based_2021}.
These approaches have been intensively pursued for a few years with great success in constitutive modeling for elastic \cite{kalina_feann_2023, kalina_automated_2022, klein_finite_2022, klein_polyconvex_2022, linden_neural_2023, henkes_physics_2022, eivazi_fe2_2023} elastoplastic \cite{masi_thermodynamics-based_2021,Vlassis2021,Vlassis2021b,Meyer2023a,Fuhg2023, rezaei_learning_2024, fuhg2023extreme, benady2024} or viscoelastic behavior \cite{tac_data-driven_2023,Abdolazizi2023a,asad_mechanics-informed_2023,Upadhyay2023a, benady2024}.
Thereby, a distinction must be made between methods with \textit{weak fulfillment} of the principles by an additional term in the loss function and \textit{strong fulfillment} with a priori compliance with the respective principle by constraining the architecture of the network \cite{cai_equivariant_2023}. According to the comparative study presented in Rosenkranz~et~al.~\cite{rosenkranz_comparative_2023}, the second approach is more promising since it is more efficient in terms of required data, robust and can extrapolate very well due to the high degree of incorporated physics, but involves some difficulties.
The challenge here is to efficiently restrict the network without loosing too much flexibility. 

For elastic and especially hyperelastic materials, many works have been published on that topic, e.g., \cite{kalina_feann_2023, kalina_automated_2022, klein_polyconvex_2022, linden_neural_2023, linka_constitutive_2021, fuhg_learning_2022, tac_data-driven_2022, asad_mechanicsinformed_2022}, among others, in which different requirements for a constitutive model are incorporated in a strong sense. Thereby, an elastic potential is approximated by using an FNN with the deformation or strain invariants as input. To allow the calibration of the network directly by tuples of stress and strain, the derivative of the potential with respect to the deformation, i.e., the network's input, is included into the loss, which is also called Sobolev training \cite{vlassis_geometric_2020,linden_neural_2023}. With easily accessible implementations for \textit{automatic differentiation} \cite{baydin2015automatic} in popular libraries like TensorFlow, PyTorch or JAX, this is no longer a major difficulty and is used in a wide variety of research areas \cite{daw_physics-guided_2017, raissi_physics-informed_2019}. Furthermore, \emph{polyconvex} NN models \cite{klein_polyconvex_2022,linden_neural_2023,Thakolkaran2022} have been formulated by using \emph{fully input convex neural networks (FICNNs)} as introduced by Amos~et~al.~\cite{amos_input_2016}. Also parametrized polyconvex models \cite{klein_parametrized_2023} have been formulated with \emph{partially input convex neural networks (PICNNs)} \cite{amos_input_2016}.

Regarding NN-based constitutive modeling of inelastic behavior with a strong physical background, a variety of works have been published meanwhile. Thereby, approaches applying the concept of \emph{internal variables} have shown to be particularily well suited.
Remarkable works on the NN-based modeling of plasticity in recent years are, for example \cite{masi_thermodynamics-based_2021,he_thermodynamically_2022,Vlassis2021,Vlassis2021b,Meyer2023a,Fuhg2023,Weber2023,Malik2021}, among others. In many of these works, however, the \emph{thermodynamic consistency} is only fulfilled in a weak sense by adding a penalty term to the loss or internal variables must be known prior to training. When using the models in multiscale problems, internal variables can be determined, e.g., by autoencoders \cite{masi_multiscale_2022,Vlassis2022}, but in real experimental setups the determination might not be practical.
Thus, there are still some challenges to overcome, in particular with regard to the fulfillment of physical conditions by construction or the provision of internal variables during training.
Only the elasto-plastic NN models \cite{Fuhg2023,Meyer2023a} a priori fulfill the second law of thermodynamics and do not require internal variables in the training data set at the same time.  
A pioneering NN-based approach to model \emph{viscoelastic} behavior is presented in \cite{huang_variational_2022}. Thereby, thermodynamic consistency is ensured by using a convex dissipation potential but internal variables are required for training. Several approaches based on a similar modeling strategy can be found \cite{asad_mechanics-informed_2023,tac_data-driven_2022,holthusen_theory_2023,rosenkranz_comparative_2023}. In contrast to \cite{huang_variational_2022}, however, no prior knowledge of the internal variable(s) is needed for training there.

\subsection{Objectives and contributions of this work}
As outlined above, NN-based constitutive models for viscoelastic problems using internal variables and accounting for fundamental physics as the second law of thermodynamics have so far received comparatively little attention. In addition, the techniques for providing internal variables during training are not satisfactory in every case, as either a high computational effort is required or flexibility and accuracy are not sufficient. Thus, within this contribution, we present a physics-augmented NN model for viscoelastic materials and an efficient training method which only requires stress and strain paths for training.
The model is built on the concept of generalized standard materials and is therefore thermodynamically consistent by construction. It consists of a free energy and a dissipation potential, which can be either expressed by the coordinates of their tensor arguments or by a suitable set of invariants. The two potentials are expressed by FICNNs/PICNNs \cite{amos_input_2016}.
For training of the NN model by paths of stress and strain, an efficient and flexible training method based on an LSTM cell is developed to automatically provide the internal variable(s) during the training process. The focus of this work is on a comprehensive benchmark test of the proposed method and a comparison with existing approaches. These include a method that obtains the internal variable by integrating the evolution equation over the entire sequence \cite{tac_data-driven_2023}, while the other method uses an auxiliary FNN for the internal variable(s) \cite{asad_mechanicsinformed_2022}.
Databases for training are generated by using a conventional nonlinear viscoelastic reference model, where 3D and 2D plane strain data with either ideal or noisy stresses are generated, respectively. The coordinate-based and the invariant-based formulation are compared and the three training methods are applied.

The article is organized as follows: After a short summary of the concept of generalized standard materials in \refs{sec:GSM}, the NN-based model is described in \refs{sec:Model}. 
In \refs{sec:Training}, three methods to calibrate the NN model are explained.
Subsequently, the presented model and the training methods are tested and compared within numerical examples in \refs{sec:Appl} using different data sets with ideal or noisy stress data. After a discussion of the results, some closing remarks are given in \refs{sec:Conclusion}.

\paragraph{Notation}
Throughout this work, several important quantities are tensors of different ranks. The space of tensors of rank $n\geq1$ is denoted as $\mathcal{L}_{n}$.
Especially, the cases of $n=2$ and $n=4$, i.e., tensors of rank two and four, are of importance herein. A tensor of rank two is denoted with upright bold symbols as \mbox{$\mathbf{T} = T_{ij} \ve{e}_i \dyad \ve{e}_j \in \tensorsofrank{2}$} and a tensor of rank four with blackboard bold symbols as \mbox{$\tte{C} = C_{ijkl} \ve{e}_i \dyad \ve{e}_j \dyad \ve{e}_k \dyad \ve{e}_l \in \tensorsofrank{4}$}. Therein, the Einstein summation convention is used, $\dyad$ is the dyadic product and \mbox{$\ve{e}_i \in \tensorsofrank{1}$} is the $i$-th Cartesian basis vector.
For tensors of rank two, only symmetric second order tensors  $\mathbf{S}\in\sym\subset\tensorsofrank{2}$ are relevant in the scope of this work, with \mbox{$\sym = \left\lbrace\,\mathbf{S}\in\tensorsofrank{2} \mid \mathbf{S}=\mathbf{S}^\top \, \right\rbrace$} and \mbox{$\mathbf{S}^\top=S_{ji}\ve{e}_i \dyad \ve{e}_j$} denoting the transpose of $\mathbf{S}$.
For tensors of rank four, the subset $\symFour\subset\tensorsofrank{4}$ of tensors with major and minor symmetries is defined as 
\mbox{$\symFour = \left\lbrace\,\tte{C}\in\tensorsofrank{4} \mid C_{ijkl}=C_{ijlk}=C_{jikl}=C_{klij} \, \right\rbrace$}. Some special fourth order tensors required here are the fully symmetric fourth order identity tensor $\IIS=\frac{1}{2}\left( \delta_{ik}\delta_{jl} + \delta_{il}\delta_{jk} \right) \allowbreak \ve{e}_i \dyad \ve{e}_j \dyad \ve{e}_k \dyad \ve{e}_l \in \symFour$, the spherical projector $\IIK=\frac{1}{3}\te{1}\dyad\te{1} \in \symFour$ with $\te{1}=\delta_{ij}\ve{e}_i\dyad\ve{e}_j$ and the deviatoric projector given by \mbox{$\IID=\IIS-\IIK$}. In the above definitions, $\dd$ denotes double contraction of adjacent indices.
Furthermore, $\tr(\mathbf{T})= T_{kk}$ is the trace of $\mathbf{T}\in\tensorsofrank{2}$, the square of a tensor is $\mathbf{T}^2=T_{ik}T_{kj}\ve{e}_i\dyad\ve{e}_j$ and the Frobenius norm is denoted by $\norm{\mathbf{T}}$ in this work and is defined as \mbox{$\norm{\mathbf{T}}=\sqrt{\mathbf{T}\dd\mathbf{T}^\top}$}.

\section{Generalized standard materials}
\label{sec:GSM}
The concept of \emph{generalized standard materials (GSMs)} \cite{miehe_incremental_2011, miehe_homogenization_2002, miehe_straindriven_2002, helmig_mathematical_2006, ziegler_derivation_1987, rice_inelastic_1971, capriz_unilateral_2011, halphen_sur_1975}   allows the formulation of thermodynamically consistent constitutive models that are entirely described by two scalar valued functions. In the context of this work, thermodynamic consistency is equivalent to satisfying the Clausius-Duhem inequality
\begin{equation}
\label{eq:Dissiaption}
    \mathcal{D} = \sig \dd \dot\eps - \dot\psi \geq 0 \quad ,
\end{equation}
where $\mathcal{D}$ is the dissipation rate, $\sig \in \sym$ is the stress, $\dot\eps \in \sym$ is the strain rate, and $\psi$ is the Helmholtz free energy density, or \emph{free energy} for short. This free energy $\psi(\eps, \iv_\alpha)$ depends on the current strain $\eps \in \sym$ and a set of internal variables $\iv_\alpha \in \sym$. The stress $\sig$ and the internal forces $\intf_\alpha \in \sym$ are obtained by differentiating the free energy with respect to the strain or the internal variables, respectively, i.e.,
\begin{equation}
    \sig = \diffp{\psi}{\eps} \quad \text{and} \quad \intf_\alpha = -\diffp{\psi}{\iv_\alpha} \quad .
\end{equation}
In the context of GSMs, there exists another potential besides the free energy, the so-called \emph{dissipation potential} $\phi(\ivdot_\alpha, \iv_\alpha, \eps)$, which depends on the rates of the internal variables $\ivdot_\alpha$ and possibly on the strain and the internal variables themselves. Differentiating the dissipation potential with respect to the rate of the internal variables again yields the internal forces
\begin{equation}
    \intfhat_\alpha = \diffp{\phi}{\ivdot_\alpha} \quad ,
\end{equation}
which are denoted with an additional $\hat{(\cdot)}$ to distinguish them from the internal forces $\intf_\alpha$ obtained with the free energy.
To construct a constitutive model that complies with Ineq.~\eqref{eq:Dissiaption}, the dissipation potential must satisfy all of the following three requirements:
\begin{enumerate}[label=(\roman*), nosep, leftmargin=0.8cm]
    \item $\phi(\ivdot_\alpha, \iv_\alpha, \eps)$ must be convex in all $\ivdot_\alpha$,
    \item $\phi(\ivdot_\alpha=\zero, \iv_\alpha, \eps) = 0$ and
    \item $\phi(\ivdot_\alpha, \iv_\alpha, \eps) \geq 0$.
\end{enumerate}
The material law is formulated with the \emph{Biot equation} \cite{biot_maurice_a_mechanics_1965}, that relates both potenials via
\begin{equation}
    \label{eq:Biot01}
    \diffp{\psi}{\iv_\alpha} + \diffp{\phi}{\ivdot_\alpha} = \te{0} \quad \text{or equivalently} \quad \intf_\alpha = \intfhat_\alpha \quad .
\end{equation}
Evaluating \refe{eq:Biot01} gives rise to the evolution laws for the internal variables. The stated conditions on $\phi$ automatically construct these evolution laws such that inequality \refe{eq:Dissiaption} is satisfied. This framework of GSMs contains various classical constitutive models, including the viscoelastic reference model in \refs{subsec:RefMat} that is used to generate the data for the numerical experiments.

\section{Formulation of the physics-augmented neural network-based model}
Based on the theoretical background from \refs{sec:GSM}, an NN-based constitutive model for viscoelastic materials is presented.
The model adapts the concept of generalized standard materials, where the two potentials are expressed as FNNs with incorporated \emph{convexity} constraints \cite{amos_input_2016, linden_neural_2023, asad_mechanics-informed_2023, klein_finite_2022, klein_parametrized_2023, fuhg_learning_2022, tac_data-driven_2023}, see App.~\ref{sec:NN}.
First, the modeling approaches for free energy and dissipation potential are described, followed by the prediction process using the adapted free energy and dissipation potential.
The training methods to find the weights and biases of the FNNs without requiring the internal variable in the training data set are explained in \refs{sec:Training}.
\label{sec:Model}
    \subsection{Modeling of the potentials}
    \label{subsec:Pot}
    The potentials $\psi$ and $\phi$ can be either expressed in terms of tensor coordinates ($\psi(\eps, \iv)$ and $\phi(\ivdot, \iv, \eps)$) or with a set of invariants of those tensors ($\psi(\invars^\psi)$ and $\phi(\invars^\phi)$) \cite{fuhg_stress_2023,holthusen_theory_2023,linden_neural_2023}.
    Since some details must be taken into account in the invariant formulation, this formulation is explained in more detail below. The coordinate formulation is obtained analogously with the corresponding simplifications.
        \subsubsection{Free energy}
        \label{subsubsec:Psi}
        The free energy is additively decomposed into an equilibrium part \mbox{$\psieq(\eps)$} and a non-equilibrium part \mbox{$\psiov(\eps, \iv)$}, i.e., $\psi(\eps, \iv) =\ \psieq(\eps)\ + \psiov(\eps, \iv)$. For the non-equilibrium part it is assumed, that the dependency on \mbox{$(\eps, \iv)$} can be expressed as $\psiov(\eps, \iv) =\ \psiov(\epsel(\eps, \iv))$, where $\epsel = \eps - \iv$ \cite{asad_mechanics-informed_2023}. The free energy is formulated with the following assumptions:
		\begin{enumerate}[label=(\roman*), nosep]
		\item $\psi$ is an isotropic tensor function.
		\item In the initial state \mbox{$\eps=\iv=\zero$}, the free energy vanishes, i.e., \mbox{$\psi(\zero, \zero)=0$}.
		\item In the initial state, the stress vanishes, i.e., \mbox{$\del{\eps}\psi(\zero, \zero)=\zero$}.
		\item In the initial state, the internal force vanishes, i.e.,\\ \mbox{$-\del{\iv}\psi(\zero, \zero)=\zero$}.
		\item $\psi(\eps, \iv)$ is assumed to be convex in $\eps$ and $\iv$.
		\end{enumerate}
        Condition (v) is used for several reasons. On the one hand, the convexity of $\psi$ results in the positive definiteness of the material tangent $\partial_{\eps\eps}\psi$, which offers numerical advantages in the application \cite{liu_generic_2020}.
On the other hand, the resulting restrictions simplify the training of the model. Furthermore, (v) is fulfilled for the used reference model from \refs{subsec:RefMat}. Furterhmore, it should be noted that other works assume the convexity of $\psi$ as well \cite{flaschel_automated_2023,asad_mechanics-informed_2023}.

		These constraints have to be incorporated into the FNN representation of $\psi$, where $\psieq$ and $\psiov$ are each described by an FNN. These FNNs are denoted as $\psieqNN$ and $\psiovNN$, respectively. Requirements (i)--(v) can be satisfied as follows:
	    \begin{enumerate}[label=(\roman*), nosep]
		\item In order to ensure an isotropic free energy, $\psieq$ and $\psiov$ are expressed in terms of invariants of their tensor arguments \cite{linden_neural_2023}, summarized in the invariant set $\invars^{\psieq} =\ ( I^{\psieq}_\alpha )_{\alpha=1}^3$ and $\invars^{\psiov}=\ ( I^{\psiov}_\alpha )_{\alpha=1}^3$, resp. That is, $\psieq(\eps)$ becomes $\psieq(\invars^{\psieq})$ and $\psiov(\epsel)$ becomes $\psiov(\invars^{\psiov})$ with
        \begin{align}
            \invars^{\psieq}(\eps) &= (\,\tr{\eps},\,  \tr{\eps^2},\, \tr{\eps^4}\,) \label{eq:InvarsPsiEq}\\
            \invars^{\psiov}(\epsel) &= (\,\tr{\epsel},\, \tr{{\epsel}^2},\, \tr{{\epsel}^4}\,) \quad . \label{eq:InvarsPsiOv}
        \end{align}
        The specific choice of the invariant sets is explained in (v), where the convexity properties are discussed.
		\item In order to ensure \mbox{$\psi(\zero,\zero)=0$}, each part $\psieq(\invars^{\psieq}(\eps))$ and $\psiov(\invars^{\psiov}(\epsel))$ is set to zero in the initial state, i.e., \mbox{$\psieq(\invars^{\psieq}(\zero))=0$} and \mbox{$\psiov(\invars^{\psiov}(\zero))=0$}. These requirements are not incorporated into the network functions themselves. Instead, a correction term is appended to the definitions of $\psieq(\invars^{\psieq})$ and $\psiov(\invars^{\psiov})$, that is, $\psieq(\invars^{\psieq})=\ \psieqNN(\invars^{\psieq}) -\  \psieqCorrPsi$ and $\psiov(\invars^{\psiov})=\ \psiovNN(\invars^{\psiov}) - \psiovCorrPsi$, where $\psieqCorrPsi=\ \psieqNN(\invars^{\psieq}(\eps=\zero))$ and $\psieqCorrPsi=\ \psieqNN(\invars^{\psiov}(\epsel=\zero))$.
		\item For the stress to vanish in the initial state, the equilibrium stress $\del{\eps}\psieq$ and the overstress $\del{\eps}\psiov$ must vanish. This is enforced with another set of correction terms $\psieqCorrSig$ and $\psiovCorrSig$, such that $\psieq(\invars^{\psieq})=\ \psieqNN(\invars^{\psieq}) -\ \psieqCorrPsi -\ \psieqCorrSig(\invars^{\psieq})$ and $\psiov(\invars^{\psiov})=\ \psiovNN(\invars^{\psiov}) -\ \psiovCorrPsi -\ \psiovCorrSig(\invars^{\psiov})$ \cite{huang_variational_2022,asad_mechanicsinformed_2022,linden_neural_2023}. These correction terms are defined such that, when differentiated with respect to $\eps$, they compensate the error in the initial stress prediction by $\psieqNN(\invars^{\psieq})$ or $\psiovNN(\invars^{\psiov})$, respectively. A possible, but inconvenient way to achieve this is to set
        \begin{equation}
            \psieqCorrSig(\invars^{\psieq}, \eps)=\sum_{\alpha=1}^{3}\left[ \evalat{\diffp{\psieqNN}{I^{\psieq}_\alpha}}{\eps=\zero} \evalat{\diffp{I^{\psieq}_\alpha}{\eps}}{\eps=\zero}\right] \dd \eps
        \end{equation}
   and analogous for $\psiov$. However, this leads to a free energy that depends on not only the invariants, but also on the strain tensor itself, which may violate (i). Therefore, $\psieqCorrSig(\invars^{\psieq})$ and $\psiovCorrSig(\invars^{\psiov})$ may only depend on the invariants. This is achieved with
            \begin{gather}
                \psieqCorrSig(\invars^{\psieq}) = \evalat{\diffp{\psieqNN}{I^{\psieq}_1}}{\eps=\zero} I^{\psieq}_1 \quad \text{and}\\
                \psiovCorrSig(\invars^{\psiov}) = \evalat{\diffp{\psiovNN}{I^{\psiov}_1}}{\epsel=\zero} I^{\psiov}_1 \; .
            \end{gather}
			Since $I^{\psieq}_1=\tr{\eps}$ and $I^{\psiov}_1=\tr{\epsel}$ are linear functions in $\eps$ and $\epsel$, this does not effect convexity of $\psi$.
		\item The internal force vanishes, if \mbox{$\evalat{-\del{\iv}\psiov}{\epsel=\zero} = \zero$}. Since \\ \mbox{$-\del{\iv}\psiov=\del{\eps}\psiov$} holds and \mbox{$\evalat{\del{\eps}\psiov}{\epsel=\zero}=\zero$} is already assured with (iii), this requirement is already secured.
		\item Convexity in $\eps$ and $\iv$ can be achieved if two conditions are met: (a) The chosen invariants are convex with respect to $\eps$ and $\iv$ and (b) $\psieqNN$ and $\psiovNN$ are convex and non-decreasing in those invariants \cite{klein_polyconvex_2022}. Since both $\psieq(\eps)$ and $\psiov(\epsel)$ depend on a single symmetric tensor of rank two, a complete set of invariants composes, e.g., the invariants $(\tr{\tensor},\, \tr{\tensor^2}, \, \tr{\tensor^3})$, where $\tensor\in\sym$. The third invariant $\tr{\tensor^3}$, however, is not convex in $\tensor$, as shown in App.~\ref{sec:Convexity}. Therefore, the functional basis is adjusted suitably by replacing $\tr{\tensor^3}$ by the convex invariant $\tr{\tensor^4}$. This is allowed, since $\tr{\tensor^3}$ can be expressed by $\tr{\tensor}$, $\tr{\tensor^2}$ and $\tr{\tensor^4}$ using the Cayley-Hamilton theorem and the original basis can be recovered. A comment on the convexity of the used invariants can be found in Appendix \ref{sec:Convexity}. Consequently, the invariant bases in Eqs. (\ref{eq:InvarsPsiEq}) and (\ref{eq:InvarsPsiOv}) allow for the construction of convex functions using \emph{non-decreasing fully input convex neural networks(FICNNs)}\footnote{Since the invariants $\tr{\eps}$ and $\tr{\epsel}$ are linear functions of $\eps$ and $\iv$, the constraints on the corresponding weights in the passthrough layers of the non-decreasing FICNN as described in \refs{sec:NN} are in general too restrictive. However, for the here shown NN-based model and the used reference material, this does not effect the prediction quality and is henceforth ignored.} \cite{amos_input_2016,klein_polyconvex_2022,linden_neural_2023,Bahmani2023} as described in App.~\ref{subsec:FICNN}.
		\end{enumerate}
		Summarizing, the free energy is modeled with an additive decomposition, where each part is a non-decreasing FICNN with additional correction terms and is expressed in a set of three convex invariants, such that
		\begin{align}
		\psi(\eps, \iv) &= \psieq(\eps) + \psiov(\epsel(\eps,\iv)) \quad , \quad \text{where}\\
		\psieq(\eps) &= \psieq(\invars^{\psieq}(\eps)) = \psieqNN(\invars^{\psieq}) - \psieqCorrPsi - \psieqCorrSig(\invars^{\psieq}) \, \text{and} \\
		\psiov(\epsel) &= \psiov(\invars^{\psiov}(\epsel)) = \psiovNN(\invars^{\psiov}) - \psiovCorrPsi - \psiovCorrSig(\invars^{\psiov}) \quad .
		\end{align}
		With that, the free energy representation is physically meaningful, even in the untrained state of the NN model. Details on the hyperparameters of the FICNNs can be found in \refs{sec:ArchitectureDetails} To simplify notation, the weights and biases of the two FICNNs are summarized in a single parameter set $\boldsymbol{\mathscr w}^\psi$.
        The dissipation potential is constructed using similar techniques.
		
        \begin{figure*}[h!]
            \centering
            \includegraphics{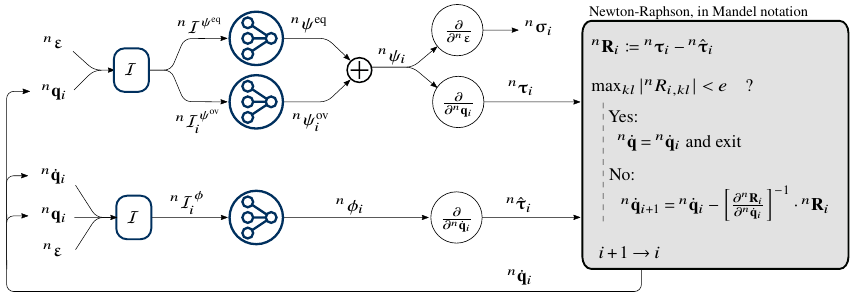}\\
            \caption{Structure of the invariant-based two-potential model and prediction process. The rate of the internal variables is determined iteratively so that the internal stress $\intf$ calculated from the free energy and the internal stress $\intfhat$ calculated from the dissipation potential are equal within a specified tolerance $e$.}
            \label{fig:Prediction}
        \end{figure*}

        \subsubsection{Dissipation potential}
        \label{subsubsec:Phi}
        The NN-based dissipation potential depends on $(\ivdot, \eps, \iv)$ \cite{rosenkranz_comparative_2023}. Similar to the non-equilibrium part of the free energy, the dependency of $\eps$ and $\iv$ is expressed in terms of $\epsel=\eps-\iv$. Consequently, $\phi = \phi(\ivdot, \epsel(\eps, \iv))$. As discussed in \refs{sec:GSM}, the dissipation potential must obey some constraints:
		\begin{enumerate}[label=(\alph*), nosep]
		\item $\phi$ is an isotropic tensor function.
		\item $\phi$ is convex with respect to its first argument, the rate of the internal variable $\ivdot$.
		\item If the internal variable does not evolve, the dissipation potential vanishes, i.e., $\phi(\zero,\epsel)=0$.
		\item If the inelastic strain does not evolve, the dissipation potential is in a minimum, i.e., $\del{\ivdot}\phi(\zero,\epsel) = \zero$.
		\end{enumerate}
		The respective network to model this function is denoted as $\phiNN$. The stated requirements are enforced with techniques similar to those used to construct the free energy.
		\begin{enumerate}[label=(\alph*), nosep]
		\item Formulating $\phi$ in invariants enforces isotropy of the resulting function. The network $\phiNN(\invars^{\phi}(\ivdot, \epsel))$ depends on the six invariants
        \begin{equation}
            \me{I}^{\phi}(\ivdot, \epsel) = (\,\tr\ivdot,\, {\tr\ivdot}^2,\, {\tr\ivdot}^4,\, \tr\epsel,\, {\tr\epsel}^2,\, {\tr\epsel}^3\,) \quad .
        \end{equation}
        Here, no mixed invariants between $\ivdot$ and $\epsel$ are used.
		\item To enforce convexity with respect to the rate of the internal variable, a \emph{non-decreasing partially input convex neural network (PICNN)} \cite{amos_input_2016,klein_parametrized_2023} is used, which is convex with respect to the three invariants $(\,\tr\ivdot,\, {\tr\ivdot}^2,\, {\tr\ivdot}^4\,)$ and not necessarily convex with respect to $(\,\tr\epsel,\, {\tr\epsel}^2,\, {\tr\epsel}^3\,)$.
		\item Analogous to the free energy, a correction term $\phiCorrPhi$ is appended to the definition of $\phi$, such that $\phi=\phiNN - \phiCorrPhi$. $\phiCorrPhi$ compensates the offset in $\phiNN$ and is defined as $\phiCorrPhi=\phiNN(\invars^{\phi}(\ivdot=\zero, \epsel))$.
		\item By adapting the stress correction term for the free energy, a correction term $\phiCorrTau$ is appended, such that $\phi = \phiNN - \phiCorrPhi - \phiCorrTau$, where $\phiCorrTau$ is
        \begin{equation}
            \phiCorrTau = \evalat{\diffp{\phiNN}{I^\phi_1}}{\ivdot=\zero} I_1^{\phi}\quad.
        \end{equation}

		\end{enumerate}
        Summarizing, the dissipation potential is modeled with a non-decreasing PICNN and additional correction terms. The dependency on $\ivdot$ is expressed with convex invariants of $\ivdot$. That is,
        \begin{align}
            \phi(\ivdot, \eps, \iv) &= \phiNN(\invars^{\phi}(\ivdot, \epsel(\eps, \iv))) - \phi_0 - \phi_{\intf} \quad . 
        \end{align}
        The NN model is thus \emph{thermodynamically consistent} and \emph{isotropic} by construction.
        All weights and biases of the PICNN with the hyperparamters from \refs{sec:ArchitectureDetails} are summarized in the parameter set $\boldsymbol{\mathscr w}^\phi$.

    \paragraph{Coordinate formulation} The previous explanations refer to the invariant formulation. For the formulation with coordinates \cite{asad_mechanics-informed_2023}, requirements (i) and (a), i.e., the isotropy of the potentials, cannot be fulfilled. In principle, the same techniques are used to comply with the remaining requirements, but instead of convex and non-decreasing NNs, only convex NNs are used, as the tensors themselves are inputs of the networks. The coordinate formulations of the additional terms for zero energy and zero stress in the initial state are briefly explained using the example of $\psieq$: The term $\psieqCorrPsi$ for zero energy in the initial state is calculated as
    \begin{equation}
        \psieqCorrPsi = \psieqNN(\eps=\zero)
    \end{equation} and the term $\psieqCorrSig$ for the stress free initial state becomes
    \begin{equation}
        \psieqCorrSig = \evalat{\diffp{\psieqNN}{\eps}}{\eps=\zero}\dd\eps \quad .
    \end{equation}
    The other additional terms for $\psiov$ and $\phi$ are calculated analogously.
    
    \subsection{Incremental predictions with the trained models}
    \label{subsec:Pred}
        Once the weights and biases of the networks for free energy and dissipation potential have been adapted, the constitutive model can be used to predict the constitutive response to a prescribed strain path consisting of multiple discrete time steps.
    	In order to solve the evolution equation numerically, an \emph{implicit Euler} scheme is applied for the temporal discretization, in which the rate of the internal variable is approximated as \mbox{${}^{n}{\ivdot}\approx({}^{n}{\iv}-{}^{n-1}{\iv})/{}^{n}{\Delta t}$}, where the superscript $n$ corresponds to the $n$-th time step of a sequence and \mbox{${}^{n}{\Delta t} = {}^{n}{t} - {}^{n-1}{t}$} is the time increment. 
    	In each time step, the rate of the internal variable is adapted iteratively, such that \refe{eq:Biot01} is fulfilled up to a prescribed tolerance. That is, the resulting internal force, which is calculated from the free energy, i.e., $\intf=-\del{\iv}\psi$, and the internal force calculated from the dissipation potential, i.e., $\intfhat=\del{\ivdot}\phi$, are supposed to be identical within a given tolerance. This iterative solution of \refe{eq:Biot01} is done using a \emph{Newton-Raphson scheme}, as depicted in \reff{fig:Prediction}.

\section{Training methods}
\label{sec:Training}
    Since the internal variable and its rate are arguments of the free energy and the dissipation potential, they must be provided in the training process to render training possible.
    In the following, however, it is assumed, that they are not present in the training data set. For data obtained through homogenization, a few approaches have been developed to enable the determination of internal variables using autoencoders \cite{masi_multiscale_2022,Vlassis2022}. The training data set $\db=(\eps(t), \sig(t), \iv(t))$ then comprises sequences of stresses, strains and internal variable(s). In real experiments, however, the determination of internal variables is in general not possible, shrinking the data set $\db=(\eps(t), \sig(t))$ to sequences of only stresses and strains.
    This section introduces three methods to address this issue. Two methods from the literature determine the internal variable by integrating the evolution equation over the entire sequence or by means of an additional auxiliary network in the form of an FNN. In the newly developed third method, the auxiliary network is replaced by an RNN.
    \subsection{Integration}
        \begin{figure*}[p]
            \centering
            \includegraphics{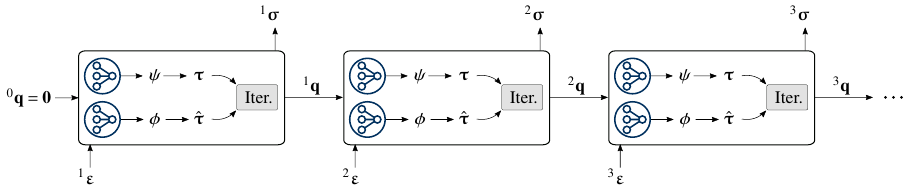}\\
            \caption{Schematic representation of the training process using the integration strategy. In each time step, the new material state is obtained iteratively with the procedure given in \reff{fig:FNN}. The internal variable is passed on to every time step as starting point for the next integration step until the last time step of the sequence is reached. Consequently, the calculation of the stress for time step $n$ requires the evaluation of all time steps $1,2,\ldots, n-1$ in advance.}
            \label{fig:Integrate}
        \end{figure*}
        The training approach denoted as integration in the following is probably the most intuitive among the three presented and has been applied in \cite{holthusen_theory_2023, tac_data-driven_2023}. In every training epoch, the stress response to individual sequences is determined following the scheme provided in \reff{fig:Prediction}. Starting from the initial state ${}^{0}\eps={}^{0}\sig={}^{0}\iv=\zero$, the stress ${}^{1}\sig$ for the new strain ${}^{1}\eps$ and the corresponding time step ${}^{1}\Delta t$ is calculated. The obtained internal variable is then passed on to the next time step and so forth until the last time step of the sequence is reached.
        The stresses $\sig$ determined this way are compared to the expected stresses $\bar\sig$ from the training data set using the loss function
        \begin{equation}
            \loss = \loss^{\sig} \quad \text{with} \quad \loss^{\sig}=\mae(\sig, \bar\sig)/s_{\sig}  \quad , \label{eq:LossInt}
        \end{equation}
        where $\mae(\sig, \bar\sig)$ is the mean absolute error between predicted stress $\sig$ and expected stress $\bar\sig$. Compared to the mean squared error, the mean absolute error performed best for the problems in this article and is therefore used within the scope of this work. Furthermore, $s_{\sig}$ is the normalization factor for the stress, that normalizes the loss to magnitude 1. The calculation of $s_{\sig}$ is explained in \ref{sec:Normalization}.
        To adapt the weights and biases of the networks, the loss function \refe{eq:LossInt} is minimized in the optimization problem
        \begin{equation}
           (\hat{\boldsymbol{\mathscr w}}^\psi,\hat{\boldsymbol{\mathscr w}}^\phi) = \underset{\boldsymbol{\mathscr w}^\psi \in \mathcal C^\psi,\boldsymbol{\mathscr w}^\phi \in \mathcal C^\phi}{\arg \min} \loss \; ,
        \end{equation}
        where the parameters $\boldsymbol{\mathscr w}^\psi$ and $\boldsymbol{\mathscr w}^\phi$ correspond to the weights and biases of the two FICNNs of the free energy and the PICNN of the dissipation potential. They are constrained by the sets $\mathcal C^\psi$ and $\mathcal C^\phi$ containing restrictions for weights and biases of these ICNNs as described in App.~\ref{sec:NN}.
        Due to the iterative solution scheme, the evaluation of the loss function is very expensive \cite{tac_data-driven_2023}. Hence, an optimizer with fast convergence is favorable. The optimizer Sequential Least Squares Programming (SLSQP) has shown to perform well for this application \cite{linden_neural_2023, rosenkranz_comparative_2023,kalina_feann_2023, kalina_neural_2024} and is therefore used in the numerical examples in \refs{sec:Appl}.
        
        \subsection{Auxiliary feedforward network}
        \begin{figure*}[p]
            \centering
            \includegraphics{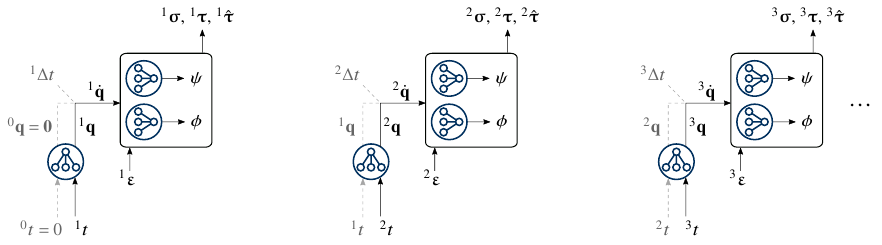}\\
            \caption{Schematic representation of the training process using an FNN as auxiliary network for the internal variable. This FNN receives a single input, the time ${}^{n}t$, and outputs the six independent entries of ${}^{n}\iv$. To calculate the rate ${}^{n}\ivdot \approx {}^{n}\iv - {}^{n-1}\iv ) / {}^{n}\Delta t$, ${}^{n-1}\iv$ is obtained by evaluating this FNN with ${}^{n-1}t$ as input.
            Note that each time step can be evaluated detached from the others, which allows fast training and the creation of minibatches within a sequence, if required.}
            \label{fig:FNN}
        \end{figure*}
        To avoid the challenges of the integration method, another method has been proposed by \citeA{asad_mechanics-informed_2023} and is adopted with slight modifications herein \cite{rosenkranz_comparative_2023}. This method introduces an additional FNN, that is supposed to learn the temporal course of the internal variable. Let $\tilde\iv(t)$ denote this additional FNN. It depends on only a single input, the time $t$, and has six outputs corresponding to the six independent entries of the symmetric tensor $\iv$.
    	For the training of the networks for free energy and dissipation potential, values for the internal variable and its rate are taken from this network, where the rate is approximated as
    	\mbox{${}^{n}\ivdot\approx({}^{n}\iv-{}^{n-1}\iv) / {}^{n}\Delta t$}
    	to be consistent with the prediction process.
    	The weights and biases of $\tilde\iv$ are adapted such that the predicted temporal course of $\tilde\iv(t)$ allows the loss function 
        \begin{align}
                &\loss = \Lsig + \Lbiot \quad \text{with} \\ &\Lsig=\mae(\sig, \bar\sig)/s_{\sig}  \quad \text{and} \quad \Lbiot=\mae(\intf, \intfhat)/s_{\sig}
        \end{align}
        to become as small as possible. The loss function consists of a term $\Lsig$ for the accurate stress prediction and another $\Lbiot$ term to comply with the Biot relation \refe{eq:Biot01} \cite{rosenkranz_comparative_2023}. Thus, the reformulated optimization problem reads 
        \begin{equation}
           (\hat{\boldsymbol{\mathscr w}}^\psi,\hat{\boldsymbol{\mathscr w}}^\phi,\hat{\boldsymbol{\mathscr w}}^q ) = \underset{\boldsymbol{\mathscr w}^\psi \in \mathcal C^\psi,\boldsymbol{\mathscr w}^\phi \in \mathcal C^\phi, \boldsymbol{\mathscr w}^q}{\arg \min} \left(\loss\right) \; ,
        \end{equation}
        where the parameter set $\boldsymbol{\mathscr w}^q$ contains weights and biases of the auxiliary FNN as specified in \refs{sec:ArchitectureDetails}.
        This leads to a reasonable representation of the internal variable without explicitly specifying its value in advance or having to integrate every sequence in every iteration of training. However, we have found that starting the optimization with randomly initialized weights for $\tilde\iv$ makes the problem too complex for common optimizers and does not lead to the desired results. In order to facilitate the training, the auxiliary network is pre-trained using the strain $\eps(t)$ as an initial guess for $\tilde\iv(t)$.
     Once the pre-training and subsequent actual training is finished, the auxiliary network is no longer necessary and the prediction process can be carried out using only free energy and dissipation potential.
     \paragraph{Remark 1}
     With this architecture, the temporal course of the internal variable for a single sequence can be modeled via a single FNN. However, if data from several sequences are available in the data set, another FNN must be added for each additional sequence, making the optimization problem increasingly difficult.
	
        \subsection{Auxiliary recurrent network}
        \begin{figure*}[p]
            \centering
            \includegraphics{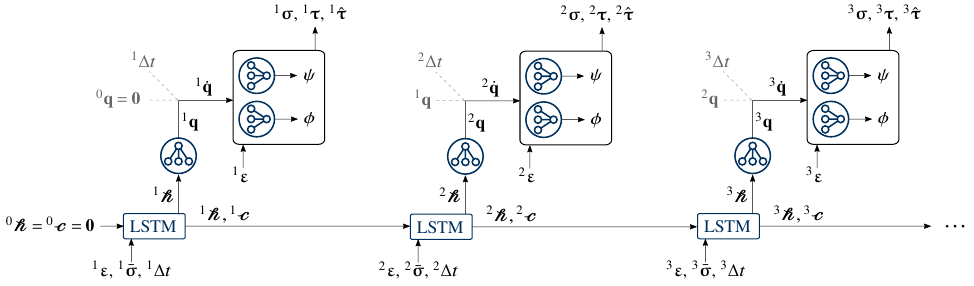}\\
            \caption{Schematic representation of the proposed training process using an RNN as auxiliary network for the internal variable. In each time step $n$, this RNN receives three inputs: the strain ${}^{n}\eps$, the stress ${}^{n}\sig$ and the time increment ${}^{n}\Delta t$. Together with the hidden state ${}^{n-1}\boldsymbol{\mathscr{h}}$ and cell state ${}^{n-1}\boldsymbol{\mathscr{c}}$ from the previous time step, the RNN-cell processes these data into a new hidden state, that contains the information about the new internal variable. This new hidden state is forwarded to an FNN, that reduces the dimensionality to six for the six independent entries of ${}^{n}\iv$. The new hidden state and cell state are passed on to the next time step, until the final time step of the sequence is reached.}
            \label{fig:RNN}
        \end{figure*}
        RNNs have proven to be particularly suitable for describing the behavior of path-dependent systems \cite{he_thermodynamically_2022,Bonatti2022,ghavamian_accelerating_2019, Heider2020}. Therefore, the auxiliary FNN is now to be replaced by an RNN.
        Using an RNN cell to generate the internal variable allows to use multiple sequences for training without requiring a new auxiliary network for every training sequence. We use an LSTM cell \cite{hochreiter_long_1997} as RNN cell. This LSTM cell has two different state vectors, that allow the cell to store information from past time steps. These state vectors are the hidden state $\boldsymbol{\mathscr{h}}$ and the cell state $\boldsymbol{\mathscr{c}}$, where $\boldsymbol{\mathscr{h}}$ is the output of the cell for every time step. The training with the auxiliary RNN works as follows: In each time step, the RNN cell receives the current strain ${}^{n}\eps$, the associated stress ${}^{n}\sig$ from the data set as well as the time step ${}^{n}\Delta t$. An FNN reduces the output of the RNN cell, the hidden state ${}^{n}\boldsymbol{\mathscr{h}}$, to 6 entries corresponding to the 6 independent entries of ${}^{n}\iv$. The rate of the internal variable ${}^{n}\ivdot$ is calculated using ${}^{n}\Delta t$ and ${}^{n-1}\iv$ from the previous time step. Through differentiation of $\psi({}^{n}\eps, {}^{n}\iv)$ and $\phi({}^{n}\ivdot, {}^{n}\eps, {}^{n}\iv)$, the stress ${}^{n}\sig$ and the internal forces ${}^{n}\intf$ and ${}^{n}\intfhat$ are obtained. The status of the LSTM cell, i.e., ${}^{n}\boldsymbol{\mathscr{h}}$ and ${}^{n}\boldsymbol{\mathscr{c}}$, is passed on to the next time step, to obtain the next material state and so on. Using the calculated stresses and internal forces, the loss function
        \begin{align}
                &\loss = \Lsig + \Lbiot \quad \text{with} \\ &\Lsig=\mae(\sig, \bar\sig)  \quad \text{and} \quad \Lbiot=\mae(\intf, \intfhat)
        \end{align}
        is evaluated. This loss function again contains a term for the error in the stress prediction and a term for compliance with the Biot equation.
        The optimization problem thus reads 
        \begin{equation}
           (\hat{\boldsymbol{\mathscr w}}^\psi,\hat{\boldsymbol{\mathscr w}}^\phi,\hat{\boldsymbol{\mathscr w}}^q ) = \underset{\boldsymbol{\mathscr w}^\psi \in \mathcal C^\psi,\boldsymbol{\mathscr w}^\phi \in \mathcal C^\phi, \boldsymbol{\mathscr w}^q}{\arg \min} \left(\loss\right) \; ,
        \end{equation}
        where the weights and biases of the auxiliary LSTM and the connected FNN  are summarized in the parameter set $\boldsymbol{\mathscr w}^q$. The concrete architecture of the LSTM and FNN can be found in \refs{sec:ArchitectureDetails}.
        Finally, after training, the RNN and the attached FNN can be discarded, leaving $\psi$ and $\phi$ as constitutive model.

\section{Numerical examples}
\label{sec:Appl}
\subsection{Data base}
    \subsubsection{Reference material}
    \label{subsec:RefMat}
\begin{figure}
        \centering
        \includegraphics{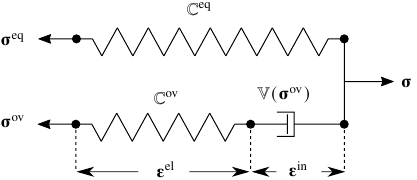}\\
        \caption{Rheological model of the viscoelastic reference solid with a single internal variable $\iv=\epsin$.}
        \label{fig:3Param}
    \end{figure}

        To generate the training data, a reference model consisting of a spring and a parallel Maxwell element as depicted in \reff{fig:3Param} is used. 
        The free energy and the dissipation potential for such a model are defined as
        \begin{align}
            \psi(\eps, \epsin) &= \frac{1}{2}\eps\dd\Ceq\dd\eps  +  \frac{1}{2}\eps^{\text{el}}\dd\Cov\dd\eps^{\text{el}} \quad \text{and}\\ 
            \phi(\dotepsin, \epsin, \eps) &= \frac{1}{2}\dotepsin\dd\Vinst(\epsin, \eps)\dd\dotepsin \quad ,
        \end{align}
        where $\dotepsin$ is defined to be the internal variable,  $\eps^{\text{el}} = \eps - \epsin$, 
        \begin{equation}
    	\label{eq:StiffnessTensors}
        	\Ceq = 3K^{\text{eq}}\IIK + 2G^{\text{eq}}\IID \quad \text{and} \quad \Cov = 3K^{\text{ov}}\IIK + 2G^{\text{ov}}\IID
    	\end{equation}
        are the equilibrium and non-equilibrium stiffness tensors, respectively, and $\Vinst(\epsin, \eps)$ is a positive semidefinite fourth order tensor describing the viscous properties of the material. This tensor is not constant, but depends on the overstress $\sigov=\Cov \dd (\eps-\epsin)$ via
        \begin{gather}
        \label{eq:ViscosityTensor}
        	\Vinst = (1-o)\Vo \exp(-\norm{\frac{1}{a}\sigov}^b) + o \quad \text{with} \\
            \Vo = 3\eta^{\text{K}}\IIK + 2\eta^{\text{D}}\IID \quad .
    	\end{gather}
     This ansatz is motivated from \cite{Kastner2012}. The specific material parameters are given in \reft{tab:MatParams}.

    \begin{table}
        \centering
        \small
        \caption{Material parameters for the viscoelastic reference material that is used to generate the training data.}
        \label{tab:MatParams}
        \centering
        \begin{tabular}{C{1.5cm} C{1.6cm} C{1.5cm} C{1.6cm}}
            \toprule
            $(\, K^{\text{eq}}$, $G^{\text{eq}}\,)$ in $\si{\mega\pascal}$ & $(\,K^{\text{ov}}, G^{\text{ov}}\,)$  in $\si{\mega\pascal}$  & $(\,\eta^\text{K}, \eta^{\text{D}}\,)$  in $\si{\mega\pascal\per\second}$  & $(\,a, b, o\,)$  in $(\,\si{\mega\pascal},\,\text{-},\,\text{-}\,)$\\
            \midrule
            $(\,500,\,300\,)$ & $(\,1000,\,700\,)$ & $(\,400,\,200\,)$ & $(\,10,\,2,\,0.1\,)$ \\
            \bottomrule
        \end{tabular}
    \end{table}

    \subsubsection{Generation of the data base}
    \begin{figure*}
        \centering
        \setlength{\figh}{3.5cm}
        \setlength{\figw}{5.5cm}
        \includegraphics{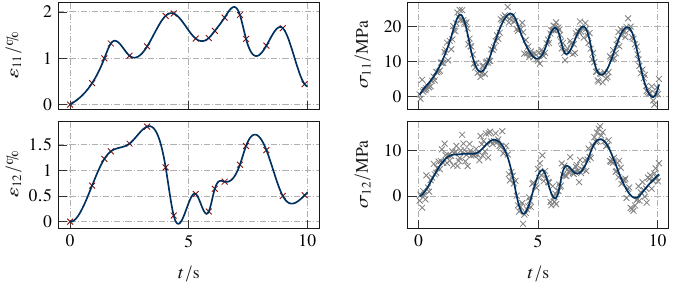}\\
        \includegraphics{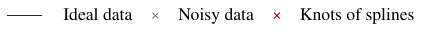}\\
        \caption{A strain path $\eps(t)$ from the training data set with respective ideal stress response $\sig(t)$ for the data sets $\db^{\text{ideal}}$ and noisy data for the data sets $\db^{\text{noisy}}$. The curve $\eps(t)$ is generated using cubic splines connecting a set of randomly sampled points, indicated as red crosses.}
        \label{fig:TrainingData}
    \end{figure*}
    
    A data set $\db=\{\,\eps(t), \sig(t)\,\}$ contains information about the temporal course of strain and stress for a single or multiple sequences.
    To generate such a sequence, the reference material model is exposed to a  randomized strain path $\eps(t)$ as shown in \reff{fig:TrainingData}. This strain path is created with cubic splines that connect a set of randomly sampled knots \cite{asad_mechanics-informed_2023,rosenkranz_comparative_2023}. Starting from $\varepsilon_{ij}^{\text{knot}}=0$ and $t^{\text{knot}}=0$, a time increment $\Delta t^{\text{knot}}$ is sampled from a uniform distribution between $\Delta t_{\text{min}}^{\text{knot}}=\SI{0.2}{\second}$ and $\Delta t_{\text{max}}^{\text{knot}}=\SI{1}{\second}$. The increments of the six independent coordinates of the strain tensor are sampled from a normal distribution with standard deviation $s_{\Delta \varepsilon}^{\text{knot}}=\SI{0.5}{\percent}$ around mean $0$. If the resulting absolute value of the strain $|\varepsilon_{ij}^{\text{knot}}+\Delta\varepsilon_{ij}^{\text{knot}}|$ exceeds $\SI{2}{\percent}$, the strain increment $\Delta \varepsilon^{\text{knot}}$ is sampled again. Once this strain path $\eps(t)$ is generated, it is applied to the reference material with random time increments $\Delta t$ from a uniform distribution between $\Delta t_{\text{min}}=\SI{0.03}{\second}$ and $\Delta t_{\text{max}}=\SI{0.07}{\second}$ to obtain the corresponding stresses $\sig(t)$.
    
    We investigate the performance of the training methods for ideal stress data $\sig$ and for noisy stress data $\Tilde{\sig}$. To receive the noisy stress data, Gaussian noise is added to the ideal stress, such that ${\Tilde\sigma}_{ij}=\sigma_{ij}+\Delta\sigma_{ij}$, where $\Delta\sigma_{ij}$ is sampled from a normal distribution with standard deviation $s^{\Delta\sigma}=\SI{1.5}{\mega\pascal}$ and mean $0$. The ideal and noisy data are shown in \reff{fig:TrainingData}.
    
    For the numerical examples, different data sets with different number of sequences and time steps per sequence are used with either ideal or noisy data. These data sets are indicated as $\db_{5\text{x}200}^{\text{ideal}}$ for a data set with 5 sequences à 200 time steps and ideal stress data, for example. To further indicate a data set that was generated using plane strain paths, i.e., with $\varepsilon_{13}=\varepsilon_{23}=\varepsilon_{33}=0$, the data set is marked with a bar as $\bar{\db}_{5\text{x}200}^{\text{ideal}}$.\\
    To investigate the performance of the trained models, they are tested on a test strain path that is generated in the same way as the sequences for the training data, that is, by sampling random knots and combining them with cubic splines. This path is scanned with constant time increment $\Delta t = \SI{0.05}{\second}$ for in total 250 time increments.

\subsection{Results}
    The presented NN-based constitutive model and training methods are now to be tested in different scenarios. In the first part of the subsection, it is exploited that the used data is generated synthetically and it is assumed that the internal variable is given in the data set to compare the invariant formulation and the coordinate formulation of the potentials in a simple test case, i.e., $\db=(\eps(t), \sig(t), \iv(t))$ additionally contains information about the internal variable. Afterwards, the internal variable is removed from the data set and the three training methods are compared for the invariant formulation, using both ideal and noisy stress data.

    \subsubsection{Training with given internal variable}
        \paragraph{Comparison invariants vs. coordinates}
            \begin{figure*}[t]
                \centering 
                \begin{subfigure}{0.4\textwidth}
                    \includegraphics[width=9cm]{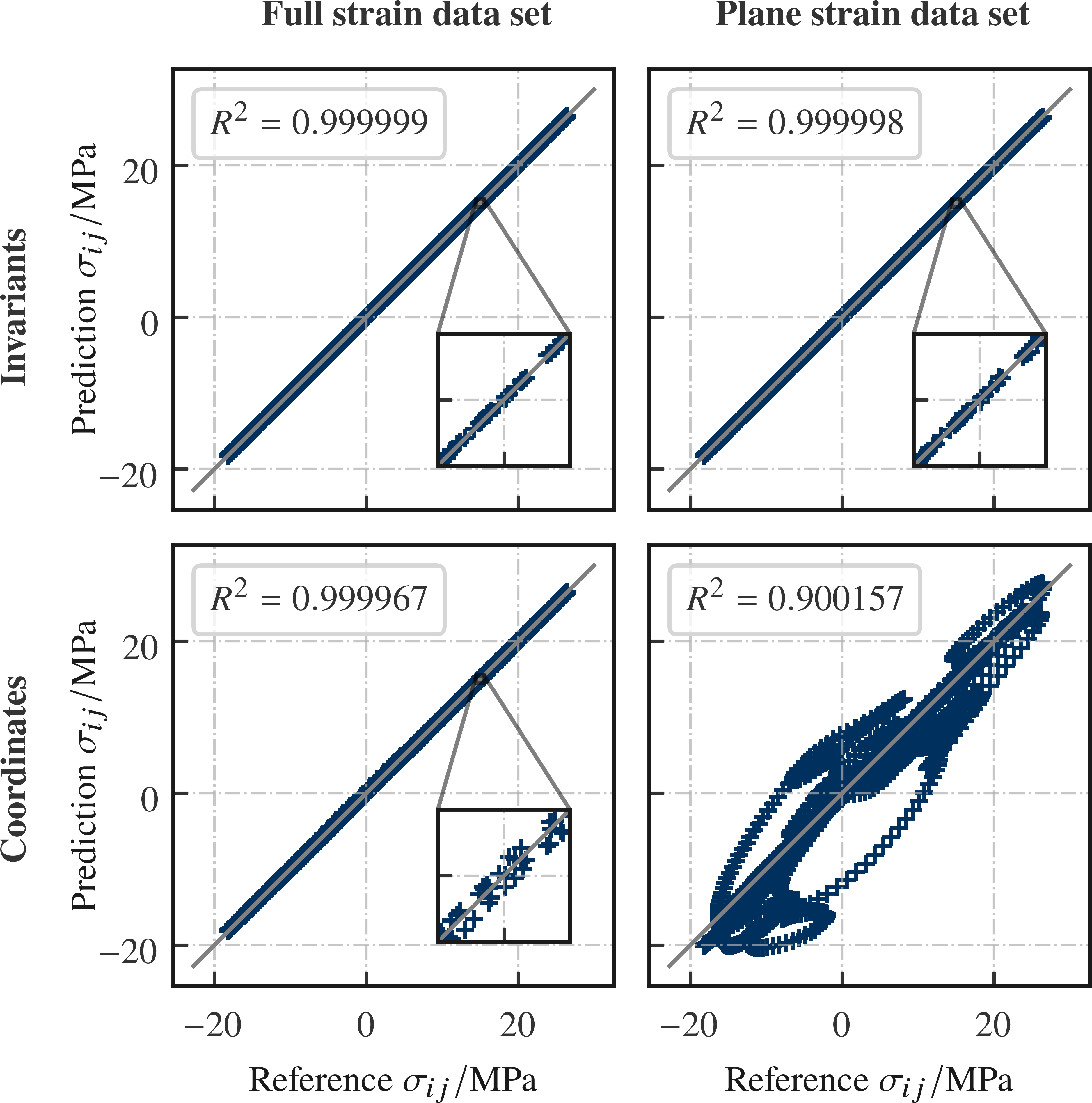}
                    \caption{}
                    \label{fig:CompInvCoord}
                \end{subfigure}
                \hfill
                \begin{subfigure}{0.4\textwidth}
                    \centering
                    \includegraphics{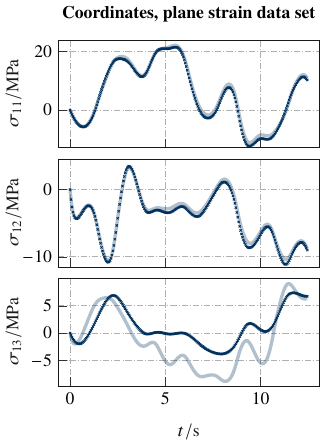}\\
                    \hspace{0.9cm}\includegraphics{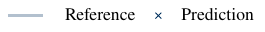}
                    \caption{}
                    \label{fig:CoordSig}
                \end{subfigure}
                \caption{Comparison of the formulation with invariants and coordinates as inputs to the networks: (a) shows correlation plots for the prediction results for both formulations using the data sets $\db_{5\text{x}200}^{\text{ideal}}$ (full strain states) and $\bar\db_{5\text{x}200}^{\text{ideal}}$ (plane strain states) with given internal variable. (b) shows the stress response $\sig(t)$ for the coordinate formulation with the data set $\bar\db_{5\text{x}200}^{\text{ideal}}$ to emphasize the lacking extrapolation capability for the out-of-plane coordinates on the example of $\sigma_{13}$.}
                \label{fig:Givenq}
            \end{figure*}
        To compare the model formulation using invariants and the formulation using tensor coordinates as inputs of the potentials, a data set $\db_{5\text{x}200}^{\text{ideal}}$ with 5 sequences à 200 time steps and the corresponding plane strain data set $\bar\db_{5\text{x}200}^{\text{ideal}}$ are used. These data sets are sufficiently large to show the full potential of both models and contain the internal variable. The training is carried out using the Adam optimizer with $20,000$ epochs, an initial learning rate of $0.01$ and an exponential learning rate decay such that the learning rate is multiplied with $0.1$ every $4,000$ epochs. To reduce the effect of the random weights initialization, both models are trained 25 times for each data set.
        The  models with the lowest final value of the loss function are then tested on the unseen test strain path. The results are shown as correlation plots in \reff{fig:CompInvCoord}.
        
        For the coordinate formulation, it can be seen that the model is able to make fairly accurate predictions for the test path based on data set $\db_{5\text{x}200}^{\text{ideal}}$. Using data set $\bar\db_{5\text{x}200}^{\text{ideal}}$, however, the model fails to extrapolate from the plane strain data to the full strain states. To illustrate this behavior, \reff{fig:CoordSig} shows the actual course of predicted stresses $\sigma_{11}$, $\sigma_{12}$ and $\sigma_{13}$. It becomes evident that the large deviations from the expected values mainly concern the out-of-plane coordinates for which the training data set lacks sufficient data.
        
        However, looking at the invariant formulation, it can be seen that it produces very accurate results regardless of the used data set, even exceeding the accuracy of the coordinate formulation with data set $\db_{5\text{x}200}^{\text{ideal}}$.
        Thus, the invariant formulation enables extrapolation from plane strain data to full strain states without notable loss of accuracy. Such a well-developed extrapolation behavior also occurs with elastic NN models based on invariants \cite{linden_neural_2023,kalina_feann_2023}. This benefit arises from the choice of a specific set of invariants that provides additional information about the anisotropy class of the underlying material law to the model. In fact, the change from plane strain to full strain states may not correspond to an extrapolation in the invariant space at all \cite{kalina_automated_2022}.

        Moreover, the invariant formulation requires significantly less training data. The decision to use 5 sequences of 200 time steps was made due to the larger amount of data needed for the coordinate formulation, whereas the invariant formulation can operate on a single sequence of 200 steps with similar prediction accuracy.
        
        For this reason, only the invariant formulation is used in the following studies and the data set is reduced to one sequence of length 200.
            
    \subsubsection{Training without given internal variable}
\begin{figure*}
            \centering
            \includegraphics[width=12cm]{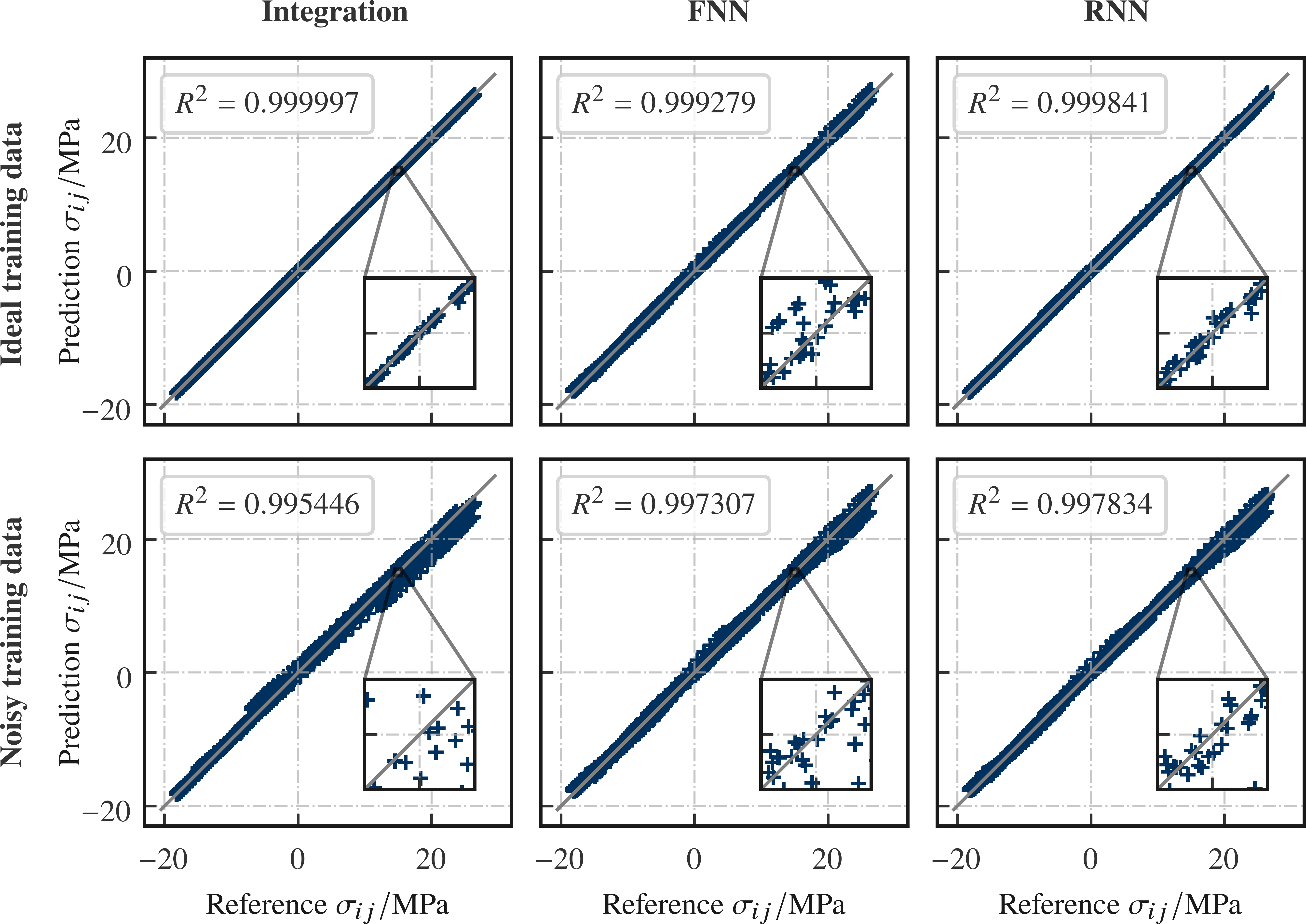}\\
            \caption{Comparison of the training methods with correlation plots for the results of the stress prediction for the unseen test path. The models were traines with the data sets $\db^{\text{ideal}}_{1\text{x}200}$ and $\db^{\text{noisy}}_{1\text{x}200}$, respectively. For the noisy data, the reference stress refers to the underlying ground truth, i.e., the noiseless data.}
            \label{fig:IntvsFNNvsRNN}
        \end{figure*}

    The internal variable is now removed from the data set and has to be generated during the training process. To do so, the methods described in \refs{sec:Training} are applied, where the data sets $\db_{1\text{x}200}^{\text{ideal}}$ and $\db_{1\text{x}200}^{\text{noisy}}$ are used to compare the methods. Again, all models are trained 25 times. The results in \reff{fig:IntvsFNNvsRNN} are generated using the model with the lowest final value of the loss function.

    \begin{table}
                \centering
                \small
                \caption{Calculation time comparison of different training methods for the model for one epoch. }
                \label{tab:calcTime}
                \centering
                \begin{tabular}{C{1.5cm} C{1cm} C{1cm} C{1cm} C{1.2cm}}
                    \toprule
                    & Integration & FNN & RNN & RNN\\
                    \midrule
                    Data set & $\db^{\text{ideal}}_{1\text{x}200}$ & $\db^{\text{ideal}}_{1\text{x}200}$ & $\db^{\text{ideal}}_{1\text{x}200}$ & $\db^{\text{ideal}}_{100\text{x}100}$\\
                    & & & &\\
                     Time per epoch & $\SI{35.0}{\second}$ & $\SI{0.021}{\second}$ & $\SI{0.059}{\second}$ & $\SI{0.112}{\second}$\\
                    \bottomrule
                \end{tabular}
        \end{table}

        \begin{figure*}
            \centering
            \includegraphics{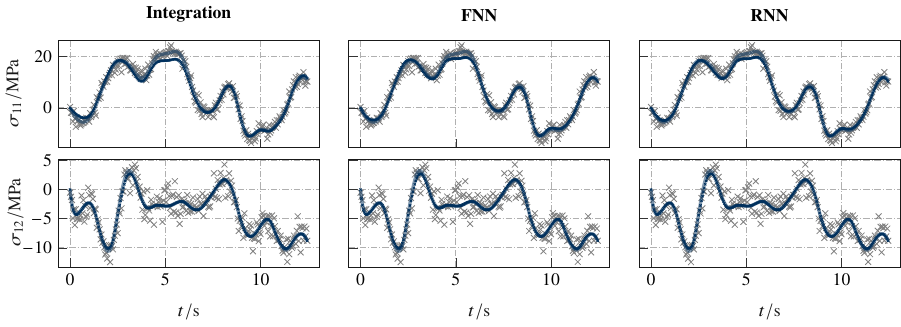}\\
            \includegraphics{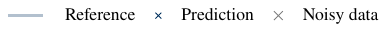}\\
            \caption{Stress responses $\sig(t)$ of the trained models for the unseen test path $\eps(t)$ on the example of the coordinates $\sigma_{11}$ and $\sigma_{12}$. The models were trained with the data set $\db^{\text{noisy}}_{1\text{x}200}$. The grey marks indicate noisy stress data to illustrate the amount of noise in the training data set.}
            \label{fig:IntvsFNNvsRNNScatter}
        \end{figure*}
        \begin{figure*}
            \centering
            \includegraphics{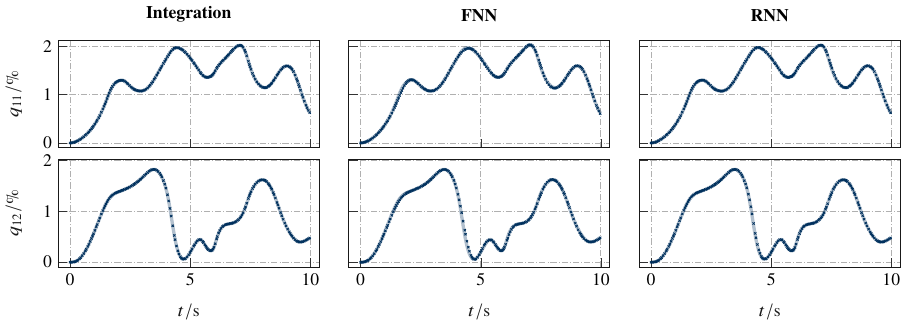}\\
            \includegraphics{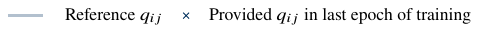}\\
            \caption{Predicted temporal course of the internal variable $\iv(t)$ of the training sequence for the three training methods using the example of the coordinates $q_{11}$ and $q_{12}$. The models were trained with the data set $\db^{\text{ideal}}_{1\text{x}200}$.}
            \label{fig:IntvsFNNvsRNNz}
        \end{figure*}
    
        \paragraph{Integration}
            \begin{figure}
                \centering
                \setlength{\figh}{4.5cm}
                \setlength{\figw}{8.4cm}
                \includegraphics{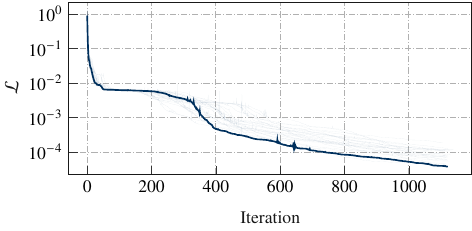}\\
                \includegraphics{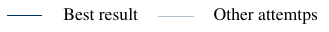}\\
            	\caption{Value of the loss function over iterations for the integration training method. The model was trained 25 times using the optimizer SLSQP, the best attempt is highlighted as thick blue curve.}
            	\label{fig:Loss_integrate}
        	\end{figure}
        The integration training method yields highly accurate results for the ideal data set and outperforms the other methods in terms of accuracy. However, it should be noted that this method uses the SLSQP optimizer for computational reasons, which has been shown to provide better results for smaller networks than Adam \cite{kalina_neural_2024}. Since no additional network is required for this method, SLSQP can be used in a computationally efficient manner with fewer function evaluations compared to Adam.
        Nevertheless, the evaluation of the loss function consumes the majority of computational time for training, since an iterative solution is required at each time step, regardless of the optimizer, see \reff{fig:Integrate}.
         Adding another sequence to the data set or doubling its length roughly doubles the cost for the evaluation of the loss function, making it more and more inefficient. However, very good results can be achieved with both the ideal training data as well as the noisy training data set as Figs.~\ref{fig:IntvsFNNvsRNN} and \ref{fig:IntvsFNNvsRNNScatter} show.

        \paragraph{Auxiliary feedforward network}
            \begin{figure}
                \centering
                \setlength{\figh}{4.5cm}
                \setlength{\figw}{8.4cm}
                \includegraphics{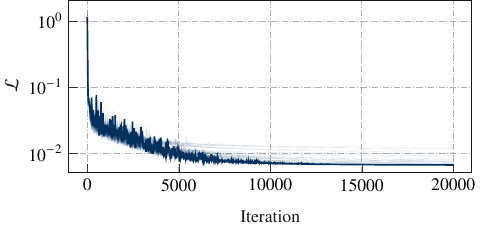}\\
                \includegraphics{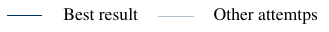}\\
        	\caption{Value of the loss function over iterations for the training method with an FNN as auxiliary network for the internal variable. The model was trained 25 times using the optimizer Adam.}
        	\label{fig:Loss_FNN}
    	  \end{figure}
       Although this training method shows the least accurate predictions for both ideal and noisy training data, the results are still sufficient in both cases. In addition to that, the computational time per iteration is very low. However, before the actual training starts, the auxiliary FNN must be pre-trained to a reasonable initial guess for $\iv(t)$ to have good starting values for the weights and biases. Using randomly initialized parameters at the beginning of the actual training did not yield useful results in the numerical experiments shown here. We used $\iv(t)=\eps(t)$ as initial guess to pre-train the FNN.
       Another major drawback of this method is the increasing demand of trainable parameters for a longer sequence or additional sequences in the training data set. Since a single FNN only predicts the temporal course of one sequence, every new sequence requires a new FNN, making the optimization more and more complex. For a single sequence of moderate length, however, the method can yield useful results, see Figs.~\ref{fig:IntvsFNNvsRNN} and \ref{fig:IntvsFNNvsRNNScatter}.
     
        \paragraph{Auxiliary recurrent network}
            \begin{figure}
                \centering
                \setlength{\figh}{4.5cm}
                \setlength{\figw}{8.4cm}
                \includegraphics{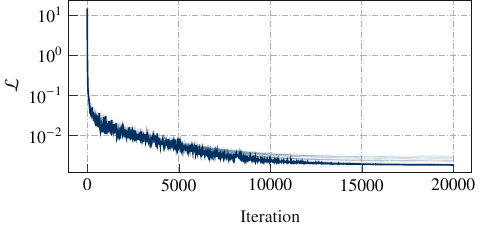}\\
                \includegraphics{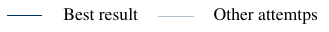}\\
        	\caption{Value of the loss function over iterations for the training method with an RNN as auxiliary network for the internal variable. The model was trained 25 times using the optimizer Adam.}
        	\label{fig:Loss_RNN}
    	\end{figure}
        \begin{figure}
            \centering 
            \includegraphics[width=5cm]{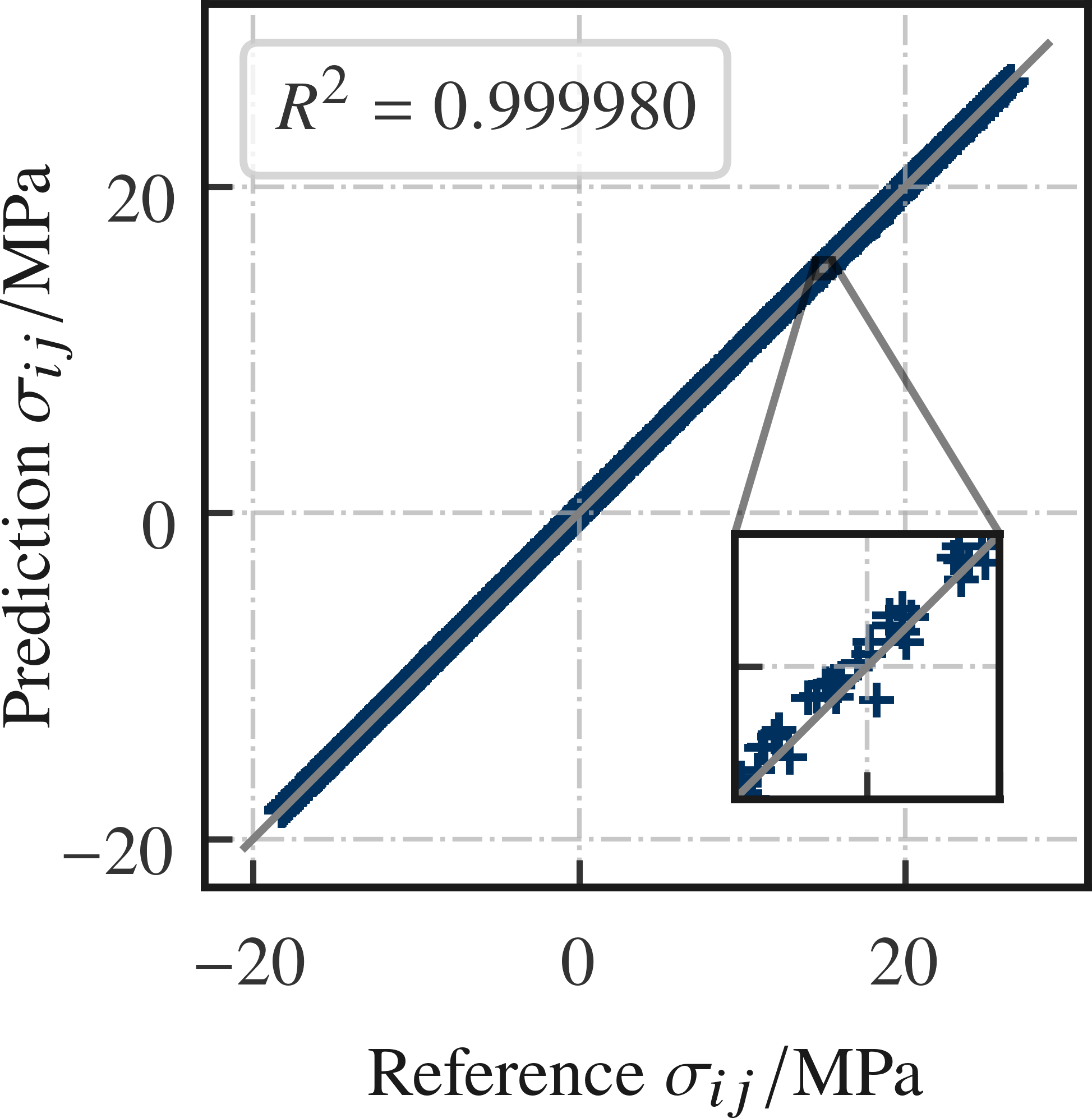}
            \caption{Correlation plot for the prediction results for the unseen test path using the RNN training method with the data set $\db_{100\text{x}100}^{\text{ideal}}$, i.e., with 100 sequences.}
            \label{fig:RNN100x100}
        \end{figure}
        The RNN as auxiliary network for the generation of the internal variable also provides precise results with both data sets. Compared to the integration method, the training is very fast and in contrast to the FNN it does not need a pre-training. 
        
        Moreover, using a training data set with several sequences instead of only one sequence does not increase the number of trainable parameters, as the RNN implicitly learns how the internal variable evolves instead of memorizing its temporal course. 
        Thus, we consider a test case which is more complex compared to the works \cite{asad_mechanics-informed_2023,holthusen_theory_2023,tac_data-driven_2023}, where only one multiaxial or even uniaxial path has been used for training. To do so, several sequences with multiaxial states are added to the training data set, i.e., $\db_{100\text{x}100}^{\text{ideal}}$ is used.
        The prediction results for the architecture trained with the set $\db_{100\text{x}100}^{\text{ideal}}$ are shown in \reff{fig:RNN100x100}. As can be seen, the stress response for the unseen strain path is even more precise, while the required time per training iteration remains short. Although the data set contains 50-times the number of time steps, the time per iteration increases from $\SI{0.059}{\second}$ to only $\SI{0.112}{\second}$. This shows that the RNN method allows to use multiple sequences efficiently without increasing amount of trainable parameters and within reasonable training time.

        \paragraph{Summary}
        Each of the analyzed techniques exhibits certain advantages compared to the others. In order to provide a comprehensive summary of this comparison, the advantages and disadvantages of each approach are listed in \reft{tab:advantages}.
        The integration training is very accurate. However, the biggest disadvantage of this method is the training time, which also affects the applicability for large data sets with many sequences, such that this combination is not practically manageable due to cost reasons.
        Using an FNN as an auxiliary network is extremely fast, but requires a pre-training of the FNN. It is also limited to one or very few sequences.
        In contrast, the RNN is fast, does not require pre-training, and can be applied to large data sets with multiple sequences without further adjustments. This feature is particularly relevant for real-world applications with experimental or homogenization data. 

        \begin{table}
            \centering
            \small
            \caption{Advantages and disadvantages of the different training methods.}
            \label{tab:advantages}
            \centering
            \begin{tabular}{C{3.0cm} C{1cm} C{1cm} C{1cm}}
                \toprule
                & Integration & FNN  & RNN \\
                \midrule
                 Good results for \\ideal data & \textcolor{myGreen}{\checkmark} & \textcolor{myGreen}{\checkmark} & \textcolor{myGreen}{\checkmark}\\
                 Good results for noisy data & \textcolor{myGreen}{\checkmark} & \textcolor{myGreen}{\checkmark} & \textcolor{myGreen}{\checkmark}\\
                 Fast training & \textcolor{myRed}{\scalebox{0.75}{\usym{2613}}} & \textcolor{myGreen}{\checkmark} & \textcolor{myGreen}{\checkmark}\\
                 No pre-training & \textcolor{myGreen}{\checkmark} & \textcolor{myRed}{\scalebox{0.75}{\usym{2613}}} & \textcolor{myGreen}{\checkmark}\\
                 Data set with many sequences & \textcolor{gray}{(\checkmark)} & \textcolor{myRed}{\scalebox{0.75}{\usym{2613}}} & \textcolor{myGreen}{\checkmark}\\
                \bottomrule
            \end{tabular}
        \end{table}

\section{Conclusions}
\label{sec:Conclusion}
This paper proposes a fast and widely applicable method to calibrate inelastic constitutive models without prior knowledge of the internal variables. This method is compared comprehensively with two existing methods. For this comparison, we propose and use a physics-augmented NN-based model for viscoelastic materials based on the concept of GSMs.

In the beginning of this work, the concept of GSMs is briefly described. Subsequently, the NN-based model is presented, which consists of two potentials: the free energy and the dissipation potential. These potentials are constrained in a physically meaningful way so that thermodynamic consistency is ensured in advance. The potentials can be expressed using either tensor coordinates or invariants of the tensor arguments. Following the presentation of the NN-based constitutive model, three training methods are introduced. These comprise a method that integrates entire sequences and two methods that use an additional NN, either an FNN or an RNN, to provide the internal variable.
Based on these techniques, numerical experiments are conducted using synthetically generated data from classical constitutive models.
First, it is assumed that the internal variable is present in the training data set. Using this data set, the invariant formulation is compared to the coordinate formulation. It is shown, that the invariant formulation requires less data, yields more accurate results and exhibits far better extrapolation capabilities when predicting stresses for arbitrary strain paths with plane strain training data. Henceforth, only the invariant formulation is used in the subsequent studies. In order to compare the three training methods, the internal variable is erased from the data set and the performance of the methods is examined using only stress and strain data. It shows, that all three methods are able to provide models that yield accurate predictions for unseen data, even for a small data set with noisy stress data. However, only the proposed RNN model allows to achieve fast and accurate results with large data sets comprising many sequences and is therefore the method with the broadest applicability.
Thus, this paper proposes a flexible constitutive model for viscoelastic materials and an efficient method for calibrating such models without prior knowledge of the internal variable. The presented training method exceeds the application possibilities of existing approaches and can form the basis for future extensions in the context of NN-based data driven constitutive modeling.

However, this study also has limitations. Specifically, the presented algorithms assume that constitutive behavior can be modeled with a single internal variable expressed by a symmetric second-order tensor. While this assumption is valid for the synthetically generated data used herein, it may not be sufficient to generalize for more complex behavior.  In this case, the number of internal variables can be increased stepwise until the desired accuracy is achieved. 
Therefore, possible future studies include the implementation and testing of such algorithms, the application to elastoplastic materials \cite{Meyer2023a,Vlassis2021}, extension to viscoelasticity at finite strains \cite{asad_mechanics-informed_2023} as well as the use of experimental \cite{Abdolazizi2023a} or homogenization data \cite{kalina_feann_2023}.

\section*{Acknowledgement}
\small
All presented computations were performed on a HPC-Cluster at the Center for Information Services and High Performance Computing (ZIH) at TU Dresden.
		The authors thus thank the ZIH for generous allocations of computer
		time. MR and MK thank the German Research Foundation (DFG) for the support within the Research Training Group GRK2868 - D$^3$. The work of KK was supported by a postdoc fellowship of the German Academic Exchange Service (DAAD) and by the Graduate Academy (GA) of TU Dresden. All supports are gratefully acknowledged.
\section*{CRediT authorship contribution statement}
\textbf{Max Rosenkranz:} Conceptualization, Formal analysis, Investigation, Methodology, Software, Validation, Visualization, Writing -- original draft, Writing -- review \& editing.
\textbf{Karl A. Kalina:} Conceptualization, Formal analysis, Investigation, Methodology, Writing -- original draft, Writing -- review \& editing
\textbf{J\"{o}rg Brummund:} Conceptualization, Formal analysis, Writing – review \& editing.
\textbf{WaiChing Sun:} Resources, Writing -- review \& editing.
\textbf{Markus Kästner:} Resources, Writing -- review \& editing.
\normalsize

\appendix
\section{Neural Networks}
\label{sec:NN}
This section describes the concept of input convex feedforward neural networks and the normalization of inputs and outputs. In order to provide a compact explanation, some symbols are introduced: The network function is denoted as $\network(\inputVec)$ and depends on an input vector $\inputVec$. The output $\outputConvPath_l$ of the layer $l$ is calculated using the activation function $\activation_l$, the biases $\biasVec_l$ and the weights $\weightMat_l$ connecting this layer to the previous layer $l-1$. The following concepts are mainly based on \citeA{amos_input_2016}, with extensions to functions that are not only convex with respect to the inputs $\inputVec$, but also with respect to some $x$ in a network $\network(\inputVec(x))$ by making the network non-decreasing in $\inputVec$ \cite{linden_neural_2023, klein_parametrized_2023}.
    \subsection{Fully input convex neural networks}
    \label{subsec:FICNN}
\begin{figure}
        \begin{subfigure}{0.4\textwidth}
        \centering
        \includegraphics{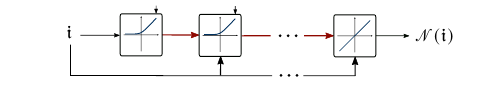}\\
        \caption{An FICNN, that is convex in $\inputVec$.}
        \label{fig:FICNNOrig}
        \end{subfigure}
    \begin{subfigure}{0.4\textwidth}
        \centering
        \includegraphics{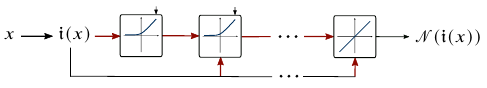}\\
        \includegraphics{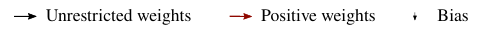}\\
        \caption{An FICNN with input $\inputVec(x)$, that is convex in $x$.}
        \label{fig:FICNNMod}
        \end{subfigure}
        \caption{Two types of FICNN architectures: the architecture in (a) is convex with respect to all entries in the input vector $\inputVec$ with unrestricted weights in the passthrough layers. The architecture in (b) is constructed, such that it is convex \textit{and} non-decreasing in $\inputVec(x)$, making it convex in $x$. This affects the weights in passthrough layers, which may no longer take negative values. This significantly limits the effectiveness of the passthrough layers, but enforces convexity in $x$. The illustrations are based on \citeA{klein_parametrized_2023}.}
        \label{fig:FICNN}
    \end{figure}

    A fully input convex neural network (FICNN) is an FNN, that is convex in all of its arguments. There are several ways to construct such an FNN, but not all variants are equivalent. Consider for example a normal FNN with non-negative weights across all layers as well as convex and non-decreasing activation functions. Such an FNN is convex, but is also extremely restrictive. Therefore, the more flexible approach of \citeA{amos_input_2016}, shown in \reff{fig:FICNN}, is adopted. It is characterized by two different sets of weights: the weights in $\weightMat_l^{\outputConvPath}$ connect the layer $l-1$ with the layer $l$ and the weights $\weightMat_l^{\inputVec}$ connect the input $\inputVec$ with the layer $l$. The outputs of the layers $l\in(1,\ldots,L)$ is now calculated as
        \begin{gather}
            \outputConvPath_{1} = \activation_1\left(\weightMat_1^{\inputVec}\inputVec + \biasVec_1 \right) \quad \text{for layer } l=1 \text{ and}\\
            \outputConvPath_{l} = \activation_l\left( \weightMat_l^{\outputConvPath}\outputConvPath_{l-1} + \weightMat_l^{\inputVec}\inputVec + \biasVec_l \right) \, \forall l \in (2,\ldots,L) \quad .
        \end{gather} 
	The network $\network(\inputVec)=\outputConvPath_{L}$ is convex if the following three conditions are fulfilled:
	\begin{enumerate}[label=(\roman*), nosep]
		\item all weights in $\weightMat_l^{\outputConvPath}$ are non-negative,
		\item all $\activation_l$ are convex, and
		\item all $\activation_l$ are non-decreasing.
	\end{enumerate}
	This approach offers greater flexibility since the weights $\weightMat_l^{\inputVec}$ remain unconstrained as shown in \reff{fig:FICNN}(a). A common choice for a convex and non-decreasing activation function is the softplus activation function, which is defined by
	\begin{equation}
	\label{eq:Softplus}
		\softplus: \R \rightarrow \RPos , \, x \mapsto \softplus(x) = \ln(1+\exp x)	 \quad .
	\end{equation}
	So far, the network has been constructed in such a way that it is convex with respect to its direct inputs. However, this must not always be sufficient. For example, in the context of this work, networks are to be implemented, that are convex with respect to a tensor, although the network inputs are invariants. The arbitrariness of the $\weightMat^{\inputVec}_l$ follows from the fact that the second derivatives of $\weightMat_l^{\inputVec} \inputVec$ with respect to $\inputVec$ vanish and consequently do not have any influence on the convexity in $\inputVec$ of the network function $\network(\inputVec)$. This changes if $\network$ is not supposed to be convex in its inputs, but in another variable $x$, which the inputs depend on. That is, $\network(\inputVec(x))$ is supposed to be convex in $x$. Then, the second derivatives of $\weightMat^{\inputVec}_l \inputVec(x)$ with respect to $x$ do not vanish if $\inputVec(x)$ is a nonlinear function of $x$. Consequently, another restriction on the weights arises, see \reff{fig:FICNN}(b): besides the conditions (i)--(iii), it must also hold that
	\begin{enumerate}[label=(\roman*), nosep]
	\setcounter{enumi}{3}
		\item all entries in $\inputVec(x)$ are convex functions of $x$, and
		\item all weights in $\weightMat^{\inputVec}_l$ are non-negative \cite{klein_polyconvex_2022}.
	\end{enumerate}

    \subsection{Partially input convex neural networks}
    \label{subsec:PICNN}
    \begin{figure*}
        \centering
        \includegraphics{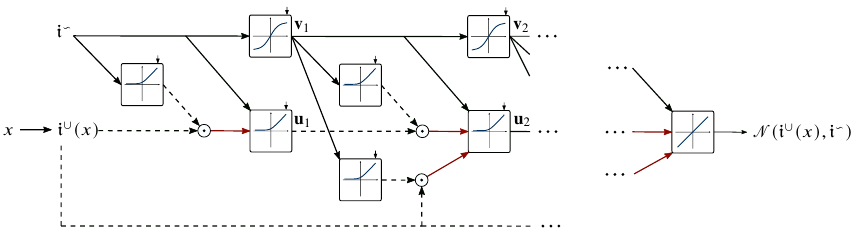}\\
        \includegraphics{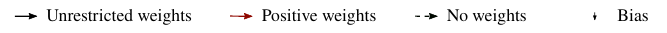}\\
        \caption{The PICNN architecture is both convex and non-decreasing in $\inputVec^{\conv}$, but allows arbitrary functional relations in $\inputVec^{\nonConv}$. For this purpose, the PICNN is divided into two paths, one for the convex part ($\outputConvPath_l$) and one for the non-convex part ($\outputNonConvPath_l$). The non-convex path $\outputNonConvPath_l$ is independent of the convex path $\outputConvPath_l$, while the layers in the convex path take into account both the output of the non-convex path as well as multiplications between $\outputNonConvPath_l$, $\outputConvPath_l$ and the convex input $\inputVec^{\conv}$. The final output is the last layer in the convex path. The illustration is based on \citeA{klein_parametrized_2023}.}
        \label{fig:PICNN}
    \end{figure*}
    A partially input convex neural network (PICNN) is, in contrast to an FICNN, only convex in some of its inputs. Let $\network(\inputVec^{\conv}, \inputVec^{\nonConv})$ denote the network, that is convex in $\inputVec^{\conv}$, but not necessarily convex in $\inputVec^{\nonConv}$. The architecture of such a network is shown in \reff{fig:PICNN}.
	The figure shows the two different paths corresponding to the convex (bottom) and non-convex part (top). The non-convex part with the layer outputs $\outputNonConvPath_l$ is a regular FNN, i.e., each layer is connected to only the previous layer of the non-convex path. A layer in the convex path with layer outputs $\outputConvPath_l$, however, depends on not only the previous layer of the convex path, but also on the output of previous layer in the non-convex path and the convex inputs. The outputs of the layers in the non-convex path are defined as
        \begin{align}
            \outputNonConvPath_{1} &= \activation^{\nonConv}_1\left( \weightMat_1^{\outputNonConvPath\outputNonConvPath}\inputVec^{\nonConv} + \biasVec_1^{\outputNonConvPath\outputNonConvPath} \right) \quad \text{for the first layer } l=1 \text{ and} \nonumber\\
            \outputNonConvPath_{l} &= \activation^{\nonConv}_l\left( \weightMat_l^{\outputNonConvPath\outputNonConvPath}\outputNonConvPath_{l-1} + \biasVec_l^{\outputNonConvPath\outputNonConvPath} \right) \, \forall l \in (2,\ldots,L-1)
        \end{align}
        and for the convex path as
        \begin{align}
        \outputConvPath_{1} = \activation^{\conv}_1 \Big(
        \weightMat_1^{\outputConvPath\inputVec^{\conv}} &\left[     \inputVec^{\conv} \hadamardProd \activation_1^{\inputVec^{\conv}\outputNonConvPath}({\tilde\weightMat}_1^{\inputVec^{\conv}\outputNonConvPath} \inputVec^{\nonConv} + \tilde\biasVec_1^{\inputVec^{\conv}\outputNonConvPath} ) \right] + \nonumber\\
        \weightMat_1^{\outputConvPath\outputNonConvPath}&\inputVec^{\nonConv} + \biasVec_1^{\outputConvPath} \Big) \quad \text{for the first layer } l=1 \text{ and}
        \end{align}
        \begin{align}
        \outputConvPath_{l} = \activation^{\conv}_l \Big( \weightMat_l^{\outputConvPath\outputConvPath} &\left[ \,    \outputConvPath_{l-1} \hadamardProd \activation_l^{\outputConvPath\outputNonConvPath}({\tilde\weightMat}_l^{\outputNonConvPath\outputNonConvPath} \outputNonConvPath_{l-1} + \tilde\biasVec_l^{\outputConvPath\outputNonConvPath} ) \right] + \nonumber\\
        \weightMat_l^{\outputConvPath\inputVec^{\conv}} &\left[     \quad \inputVec^{\conv} \hadamardProd \activation_l^{\inputVec^{\conv}\outputNonConvPath}({\tilde\weightMat}_l^{\inputVec^{\conv}\outputNonConvPath} \outputNonConvPath_{l-1} + \tilde\biasVec_l^{\inputVec^{\conv}\outputNonConvPath} ) \right] + \nonumber\\
        \weightMat_l^{\outputConvPath\outputNonConvPath}&\outputNonConvPath_{l-1} + \biasVec_l^{\outputConvPath} \Big) \quad \forall l \in (2,\ldots,L) \quad .
        \end{align}
        The final output of the network is the last layer of the convex path, i.e., $\network(\inputVec^{\conv}, \inputVec^{\nonConv})=\outputConvPath_L$.
		Instead of two sets of weights, a PICNN comprises six different sets of weights ($\weightMat^{\outputNonConvPath\outputNonConvPath}_l$, $\weightMat^{\outputConvPath\outputConvPath}_l$, $\weightMat^{\outputConvPath\inputVec^{\conv}}_l$, $\weightMat^{\outputConvPath\outputNonConvPath}_l$, $\tilde\weightMat^{\outputNonConvPath\outputNonConvPath}_l$, $\tilde\weightMat^{\inputVec^{\conv}\outputNonConvPath}_l$) to take into account possible products between the two paths and the convex input. The network is convex in $\inputVec^{\conv}$ if the following four conditions are fulfilled:
	\begin{enumerate}[label=(\roman*), nosep]
		\setcounter{enumi}{0}
		\item all weights in $\weightMat_l^{\outputConvPath\outputConvPath}$ are non-negative,
		\item all activations $\activation_l^{\conv}$ are convex,
		\item all activations $\activation_l^{\conv}$ are non-decreasing, and 
		\item all activations $\activation_l^{\outputConvPath\outputNonConvPath}$ map to non-negative values.
	\end{enumerate}		
	All other weights may as well take negative values and the activations $\activation_l^{\nonConv}$ and $\activation_l^{\inputVec^{\conv}\outputNonConvPath}$ can be chosen arbitrarily. A valid choice for $\activation_l^{\conv}$ is, like for the FICNN, the softplus function $\softplus$, which additionally is non-negative and thus can be used as $\activation_l^{\outputConvPath\outputNonConvPath}$ as well.
	The mentioned conditions yield a network $\network(\inputVec^{\conv}, \inputVec^{\nonConv})$, that is convex in all entries of $\inputVec^{\conv}$. As already discussed for the FICNN, this is not always sufficient. If a network $\network(\inputVec^{\conv}(x), \inputVec^{\nonConv})$ is supposed to be convex in $x$, it is not the first function acting on $x$, which leads to further restrictions on the weights and activations. $\network(\inputVec^{\conv}(x), \inputVec^{\nonConv})$ is convex in $x$, if in addition to (i)--(iv) it holds:
	\begin{enumerate}[label=(\roman*), nosep]
		\setcounter{enumi}{4}
		\item all entries of $\inputVec^{\conv}$ are convex functions of $x$,
		\item all  weights in $\weightMat_l^{\outputConvPath\inputVec^{\conv}}$ are non-negative, and
		\item all $\activation_l^{\inputVec^{\conv}\outputNonConvPath}$ map to non-negative values.
	\end{enumerate}
	These additional restrictions are only necessary if the entries of $\inputVec^{\conv}$ are non-linear function in $x$, i.e., the second derivatives do not vanish. Otherwise, the standard architecture is also valid.

\subsection{Normalization}
\label{sec:Normalization}
The training process of neural networks is typically more stable and efficient if the inputs and outputs of the network are values of magnitude 1. In particular, since the free energy and the dissipation potential take on large numerical values in the range \mbox{$>10^6$}, normalization is essential for a successful training. 
Normalization of a known quantity $\arbsymb$ to the range $(-1, 1)\ni \tilde\arbsymb$ can be carried out with
\begin{gather}
\tilde{\arbsymb} = \frac{\arbsymb - m_{\arbsymb}}{s_{\arbsymb}} \quad \text{with} \label{eq:Normalization}\\
s_{\arbsymb} = \frac{1}{2}(\arbsymb_{\text{max}}-\arbsymb_{\text{min}}) \quad \text{and} \\ m_{\arbsymb} = \frac{1}{2}(\arbsymb_{\text{max}}+\arbsymb_{\text{min}}) \quad ,
\end{gather}
where $\tilde{\arbsymb}$ represents the normalized quantity and $\arbsymb_{\text{max}}$ and $\arbsymb_{\text{min}}$ are the maximum and minimum value of $\arbsymb$ across the whole training data set.
For the problem described here, the inputs $\eps$, $\iv$, $\ivdot$, $t$ and $\Delta t$ or the respective invariants $\invars^{\psieq}, \invars^{\psiov}$ and $\invars^{\phi}$ have to be normalized, as well as the outputs $\psieq$, $\psiov$ and $\phi$.
The difficulty for the proposed model arises from the fact that neither $\iv$ and $\ivdot$ as inputs nor $\psieq$, $\psiov$ or $\phi$ as outputs are known before the training, but only $\eps(t)$, $\sig(t)$ and $t$ are known in advance.
For this reason, the order of magnitude of the expected values for the unknown variables is estimated based on the available variables. Therefore, the normalization parameters $s$ and $m$ are determined on the basis of the following assumptions: the internal variable and the variable $\epsel=\eps-\iv$ are normalized with the values of $\eps$, i.e., $s_{\iv}=s_{\epsel}=s_{\eps}$, while the rate $\ivdot$ is normalized with $\dot\eps$.
The normalization parameters for the potentials are defined to be $s_{\psieq}=s_{\psiov}=m_{\psieq}=m_{\psiov}=s_{\eps}\cdot s_{\sig}$, where the choice of the $m$s exploits the fact that $\psieq,\psiov,\phi \geq 0$.
Using these normalization parameters, the network itself maps only from values of magnitudes around 1 to values of magnitude 1, which enables an efficient training.
Note, that due to the linear nature of the transformation \refe{eq:Normalization}, this type of normalization does not effect the convexity properties of the resulting network function.

\section{Convexity of the invariant basis}
\label{sec:Convexity}
To enforce convexity of a network with respect to a symmetric second order tensor $\tensor\in\sym$, the invariant basis $\invars = (\,\tr\tensor, \tr\tensor^2, \tr\tensor^4 \, )$ is used throughout this work. To motivate this choice, the convexity of
\begin{enumerate}[label=(\roman*), nosep]
    \item $f(\tensor)=\tr\tensor$,
    \item $f(\tensor)=\tr\tensor^2$,
    \item $f(\tensor)=\tr\tensor^3$ and
    \item $f(\tensor)=\tr\tensor^4$
\end{enumerate}
is examined.
Two equivalent definitions of convexity are used here. A function $f(\tensor)$ is convex, if
\begin{align}
\label{eq:ConvexityFunc01}
f(\AT+\lambda\left[\BT-\AT\right]) \leq f(\AT) + &\lambda\left[f(\BT)-f(\AT)\right] \nonumber \\
&\forall \AT, \BT \in \sym , \, \forall \lambda\in[0,1] \quad .
\end{align}
or, under the assumption, that $f$ is twice continuously differentiable,
\begin{equation}
\label{eq:ConvexityFunc03}
\AT \dd \frac{\partial^2 f}{\partial \tensor \partial \tensor} \dd \AT \geq 0 \, \forall \AT \in \sym \quad .
\end{equation}
Note that for twice differentiable $f$, \refi{eq:ConvexityFunc03} follows from \refi{eq:ConvexityFunc01} and vice versa.
\begin{enumerate}[label=(\roman*), nosep]
\item From \refi{eq:ConvexityFunc01} follows directly
\begin{equation}
\label{eq:ProofJ1}
\tr(\AT + \lambda\left[\BT-\AT\right]) = \tr(\AT) + \lambda\left[\tr(\BT)-\tr(\AT)\right] \quad .
\end{equation}
As a linear function, the invariant $\tr(\tensor)$ is thus (weakly) convex.
\item Applying \refi{eq:ConvexityFunc01} to $f(\tensor)=\tr\tensor^2$ yields
\begin{align}
\label{eq:ProofJ2}
\tr\left(\left(\AT + \lambda\left[\BT-\AT\right]\right)^2 \right) = &\tr \left(\AT^2\right. + \lambda\AT\cdot\left[\BT-\AT\right] + \nonumber\\
& \lambda\left[\BT-\AT\right]\cdot\AT + \nonumber \\
& \lambda^2\left.\left[\BT^2-\AT\cdot\BT-\BT\cdot\AT+\AT^2\right]\right) \, .
\end{align}
With $\tr(\AT\cdot\BT)=\tr(\BT\cdot\AT)$, this expression simplifies to
\begin{align}
\tr(\AT^2) + 2\lambda\tr(\AT\cdot\BT-\AT^2) &+ \nonumber\\
\lambda^2\tr(\BT-2\AT\cdot\BT+\AT^2) & \leq \tr(\AT^2) + \nonumber\\ 
& \quad \lambda \left[\tr(\BT^2) - \tr(\AT^2)\right] \nonumber\\
\lambda^2\tr(\BT^2-2\AT\cdot\BT+\AT^2) &\leq \lambda \tr(\BT^2 -2\AT\cdot\BT +\AT^2) \nonumber\\
\lambda^2\tr((\BT-\AT)^2) &\leq \lambda\tr((\BT-\AT)^2) \nonumber\\
\lambda &\leq 1 \quad .
\end{align}
\item For $f(\tensor)=\tr\tensor^3$, the tensors $\AT=\zero$ and $\BT=-\one$ are considered. With \refi{eq:ConvexityFunc01} it follows
\begin{align}
\label{eq:ProofJ3}
\tr\left(\left(\zero + \lambda\left[-\one-\zero\right]\right)^3\right) = -3\lambda^3 &\leq \tr(\zero^3) + \nonumber\\
& \quad \lambda\left[\tr((-\one)^3) - \tr(\zero^3)\right] \nonumber \\
-3\lambda^3 &\leq -3\lambda \nonumber \\
\lambda^3 &\geq \lambda \quad ,
\end{align}
which is a contradiction for $\lambda\in [0,1]$. The  invariant $\tr\tensor^3$ is thus not convex in $\tensor$. 
\item To show the convexity of $f(\tensor)=\tr\tensor^4$, \refi{eq:ConvexityFunc03} is used. Evaluating \refi{eq:ConvexityFunc03} and introducing the abbreviation $A_{ik}S_{kj}=r_{ij}$ yields
\begin{equation}
    \frac{\partial f}{\partial S_{ij} \partial S_{kl}} A_{ij} A_{kl} = 4 \left[ 2r_{mn}r_{mn} + r_{mn}r_{nm} \right] \quad .
\end{equation}
Decomposing $\te{r}$ into the symmetric part $s_{ij} = s_{ji}$ and antisymmetric part $a_{ij}=-a_{ji}$, such that $r_{mn} = s_{mn} + a_{mn}$ and using $s_{mn}a_{mn} = 0$ finally leads to
\begin{equation}
    \frac{\partial f}{\partial S_{ij} \partial S_{kl}} A_{ij} A_{kl} = 4 \left[ 3 s_{mn}s_{mn} + a_{kn}a_{kn} \right] \geq 0  \quad .
\end{equation}
\end{enumerate}
Consequently, $\invars = (\,\tr\tensor, \tr\tensor^2, \tr\tensor^4 \, )$ forms a complete and convex set of invariants.

\section{Architecture details}
\label{sec:ArchitectureDetails}
\begin{table*}
    \centering
    \small
    \caption{Hyperparameters, that were used for the NNs of the potentials.}
    \label{tab:HyperParams}
    \centering
    \begin{tabular}{c | c c c c c c}
        \toprule
        Network & $\psieqNN$ & $\psieqNN$ & $\psiovNN$ & $\psiovNN$ & $\phiNN$ & $\phiNN$\\
        Input type & Invariants & Coordinates & Invariants & Coordinates & Invariants & Coordinates \\ 
        \midrule
        Architecture                & FICNN & FICNN & FICNN & FICNN & PICNN & PICNN \\
        Inputs (convex path)        & 3 & 6 & 3 & 6 & 3 & 6\\
        Inputs (non-convex path)    & - & - & - & - & 3 & 6\\
        Hidden layers               & 1 & 1 & 1 & 1 & 1   & 1\\
        Neurons in hidden layer (convex path)     & 10 & 20 & 10 & 20 & 10 & 20\\
        Neurons in hidden layer (non-convex path) & - & - & - & - & 10 & 20\\
        Outputs                     & 1 & 1 & 1 & 1 & 1 & 1\\
        Activation (convex path)     & $\softplus$ & $\softplus$ & $\softplus$ & $\softplus$ & - & - \\
        Activation (non-convex path) & - & - & - & - & $\tanh$ & $\tanh$ \\
        \bottomrule
    \end{tabular}
\end{table*}

\subsection{Potentials}
The free energy and the dissipation potential are implemented as two FICNNs and a PICNN. The details about the size of the networks, i.e., number of hidden layers, number of neurons or used activation functions can be found in \reft{tab:HyperParams}.
\subsection{Auxiliary networks}
\paragraph{Integration}
No auxiliary network is necessary for this training method.
\paragraph{FNN}
The auxiliary network for the internal variable is an ordinary FNN with a single input (the time $t$), 2 hidden layers with 50 neurons each, $\tanh$ activations in the hidden layers and 6 output neurons with linear activation. 
\paragraph{RNN}
The auxiliary network comprises an RNN cell and a subsequent FNN. The RNN cell is an LSTM cell with 13 inputs (6 for the strain tensor $\eps$, 6 for the stress tensor $\sig$ and one for the time increment $\Delta t$), 50 entries in the hidden state and $\tanh$ activations. The FNN takes the hidden state as input, has no hidden layers, i.e., it consists of only the input and output layer, and 6 has output neurons with linear activations.

\bibliographystyle{unsrtnat} 
\bibliography{references}

\end{document}